\documentclass[12pt,preprint]{aastex}

\usepackage{natbib}
\usepackage{color} 
\usepackage{ulem}  
\usepackage{psfrag}

\bibpunct{(}{)}{;}{a}{}{,}

\slugcomment{Accepted for publication in ApJ -- February 21, 2008}

\shorttitle{Azimuthal variations of the CR ion acceleration at the BW of SN 1006} 

\shortauthors{Cassam-Chena{\"\i} et al.}

\begin{document}

\title{Morphological evidence for azimuthal variations of the cosmic ray ion
acceleration at the blast wave of SN 1006}

\author{\mbox{Gamil Cassam-Chena{\"\i}\altaffilmark{1}, John P.
Hughes\altaffilmark{1}, Estela M. Reynoso\altaffilmark{2,3}, 
Carles Badenes\altaffilmark{4}, David Moffett\altaffilmark{5}}}

\altaffiltext{1}{Department of Physics and Astronomy, Rutgers
University, 136 Frelinghuysen Road, Piscataway, NJ 08854-8019;
\texttt{chenai@physics.rutgers.edu, jph@physics.rutgers.edu}} 
\altaffiltext{2}{Instituto de Astronom\'ia y F\'isica del
Espacio (IAFE), CC 67, Sucursal 28, 1428 Buenos Aires, Argentina}
\altaffiltext{3}{Departamento de F\'isica, Facultad de Ciencias Exactas y
Naturales, Universidad de Buenos Aires, Argentina}
\altaffiltext{4}{Department of Astrophysical Sciences, Princeton University,
Peyton Hall, Ivvy lane, Princeton, NJ 08544-1001}
\altaffiltext{5}{Department of Physics, Furman University, 
3300 Poinsett Highway, Greenville, SC 29613} 

\begin{abstract}
Using radio, X-ray and optical observations, we present evidence for
morphological changes due to efficient cosmic ray ion
acceleration in the structure of the southeastern region of the
supernova remnant SN 1006. SN 1006 has an apparent bipolar morphology
in both the radio and high-energy X-ray synchrotron emission. In the
optical, the shock front is clearly traced by a filament of Balmer
emission in the southeast. This optical emission enables us to trace
the location of the blast wave (BW) even in places where the synchrotron
emission from relativistic electrons is either absent or too weak to
detect. The contact discontinuity (CD) is traced using images in the
low-energy X-rays (oxygen band) which we argue reveals the
distribution of shocked ejecta. We interpret the azimuthal variations
of the ratio of radii between the BW and CD plus the X-ray and radio
synchrotron emission at the BW using CR-modified hydrodynamic models.
We assumed different azimuthal profiles for the injection rate of
particles into the acceleration process, magnetic field and level of
turbulence.  We found that the observations are consistent with a
model in which these parameters are all azimuthally varying, being
largest in the brightest regions.
\end{abstract}

\keywords{ acceleration of particles --- cosmic rays --- shock waves
--- ISM: individual (SN 1006) --- supernova remnants }

\section{Introduction\label{sect-intro}}

Collisionless shocks in young supernova remnants (SNRs) are thought to
be responsible for the production and acceleration of the bulk of
Galactic cosmic rays (CRs)  at least up to the ``knee'' ($\sim 3$ PeV)
of the CR spectrum \citep[e.g.,][]{bev07}. The theoretical mechanism
is believed to be the first order Fermi acceleration, also known as
diffusive shock acceleration (DSA), where electrons, protons and other
ions scatter back and forth on magnetic fluctuations or Alfv\'en wave
turbulence through the velocity discontinuity associated with the
shock \citep[][and references therein]{joe91, mad01}.  The presence of
the turbulence is a key ingredient in the process.  The higher the
turbulence level, the higher the scattering rate and the faster the
acceleration proceeds, allowing particles to reach higher energies.
The fast acceleration implies also that the CR pressure becomes
significant very quickly so that the shock structure is highly
nonlinear \citep[e.g.,][]{elb00, ba07_NL}. The back-reaction of the
particles on the shock hydrodynamics increases the level of magnetic
turbulence and the injection rate in turn, i.e., the accelerated
particles create the scattering environment by themselves
\citep{be78a, blo78, vob03}. Moreover, the production of relativistic
particles (extracted from the thermal plasma) along with their escape
from the shock system increases the overall density compression ratio
\citep{bee99, bl02}.

This theoretical picture leads to several observational consequences.
The intensity of the synchrotron emission from shock-accelerated
electrons will be larger as the acceleration becomes more efficient
since both the number of relativistic particles and intensity of the
magnetic field must increase. In the region of very efficient
acceleration, the X-ray synchrotron emission from the highest energy
electrons will be concentrated in the form of thin sheets behind the
blast wave due to synchrotron cooling in the high turbulent magnetic
field amplified at the shock whereas the radio synchrotron emission
will be much broader in comparison \citep{elc05}. Finally, the
morphology of the interaction region between the blast wave (BW) and
the ejecta interface or contact discontinuity (CD) will become much
thinner geometrically as the acceleration becomes more efficient due
to the change in the compressibility of the plasma \citep{dee00,
eld04}. Because shocks are believed to put far more energy into ions
than electrons, this last point, if observed, would provide direct
evidence for the acceleration of CR ions at the BW.

The detection of $\gamma$-rays with a neutral pion-decay signature
resulting from the interaction of shock-accelerated protons with the
ambient matter would provide evidence for CR ion acceleration as well.
The best chance to see a clear pion-decay signal is when a SNR
interacts with a dense medium, the prototype case being RX
J1713.7--3946 \citep{cad04b}. Very high energy $\gamma$-rays have been
detected from this remnant \citep{aha07_RXJ1713} and also from a few
others \citep[e.g.,][]{aha07_RXJ0852,hol07_RCW86}. However, the
pion-decay signal is generally difficult to extract from these
observations because there are other processes efficiently producing
$\gamma$-rays but involving CR electrons, namely bremsstrahlung and
Inverse-Compton scattering off the 2.7 K cosmic microwave background
and Galactic photons. The nature of the underlying process leading to
$\gamma$-rays and therefore the nature of the emissive particles is
subject to intense debate \citep[e.g.,][]{bev06_RXJ1713, pom06,
uca07, pl08}. Besides, in a large number of SNRs where bright and
geometrically thin X-ray synchrotron-emitting rims are observed (e.g.,
Keper, Tycho, SN 1006), ambient densities are often very low leaving
open the question of CR ion production in these SNRs \citep[see
also][]{kaw08}.

The idea that the gap between the BW and CD allows quantifying the
efficiency of CR ion acceleration was proposed by \cite{de05} in the
case of Tycho, Kepler and Cas A. In Tycho, a remnant where thin X-ray
synchrotron-emitting rims are observed all around with little
variations in brightness, \citet{wah05} showed that the BW and CD are
so close to each other that they cannot be described by standard
hydrodynamical models (i.e., with no CR component), hence providing
evidence for efficient CR ion acceleration in this SNR.  In a later
study, \citet{cah07} showed that the X-ray and to some extent the
radio properties of the synchrotron emitting rim as well as the gap
between the CD and BW were consistent with efficient CR ion
acceleration modifying the hydrodynamical evolution of Tycho.

Ideally, one would like to find a SNR where both efficient and
inefficient particle acceleration take place so that differential
measurements can be made. This offers a great advantage for the
astrophysical interpretation since a number of uncertainties (e.g.,
distance and age) can be eliminated in the comparison.  Furthermore,
by directly observing a difference in the gap between the BW and CD in
regions of efficient and inefficient acceleration, one obtains a
direct, model-independent, confirmation that CRs do indeed modify the
hydrodynamical evolution of the BW. SN 1006 provides a valuable
laboratory in that regard since the gradual variations of the
intensity of the radio and X-ray synchrotron emission along the BW
point to underlying variation of CR ion acceleration.  The task
of measuring the ratio of radii between the BW and CD in
the regions of inefficient and efficient CR acceleration will be
challenging. Nonetheless, SN 1006 is one of the rare objects (perhaps
even the unique one) where this can be accomplished as we explain below.

In this paper, we focus our attention on the poorly studied
southeastern (SE) quadrant of SN 1006, i.e., along the shock front
lying between the northeastern and the southwestern
synchrotron-emitting caps. Our first goal is to measure the ratio of
radii between the BW and CD there. We can fairly easily follow the BW
in the bright synchrotron-emitting caps (presumably the regions of
efficient CR acceleration) and the BW can still be traced from the
H$\alpha$ emission in the region where both the radio and X-ray
synchrotron emissions are either too faint to be detected or absent
(presumably the regions of inefficient CR acceleration). As for the
CD, it can be traced from the thermal X-ray emission associated with
the shocked ejecta. Our second objective is to measure the azimuthal
variations of the synchrotron brightness in the radio and several
X-ray energy bands.  With these observational key results in hand, we
use CR-modified hydrodynamic models of SNR evolution to constrain the
azimuthal variations of the acceleration parameters, i.e., the
injection rate of particles in the acceleration process, strength of
the magnetic field, particle diffusion coefficient and maximum energy
of electrons and ions.

This paper is organized as followed. In \S \ref{sect-sn1006}, we
describe the basic characteristics of SN 1006. In \S \ref{sect-data},
we present the data used in our study. In \S \ref{sect-results}, we
present the key observational results in SN 1006. In \S
\ref{sect-models-vs_obs}, we try to relate hydrodynamical models with
those results in the context of CR-unmodified and -modified shocks. 
Finally, we discuss our results (\S \ref{sect-discussion}) and present
our conclusion (\S \ref{sect-conclusion}).

\section{Basic characteristics of SN 1006\label{sect-sn1006}}

SN 1006 was a thermonuclear supernova (SN) widely seen on Earth in the
year 1006 AD. More than a thousand year later, the remnant from this
explosion is a huge shell of $30\arcmin$ angular size.

The synchrotron emission is detected from the radio up to the X-rays
and dominates in two bright limbs -- the northeast (NE) and southwest
(SW) -- where several thin ($20\arcsec$ width or so) rims/arcs are
running at the periphery, sometimes crossing each other \citep{bay03,
lor03, rob04}. Those apparent ripples are most likely the result of
the projection of undulating sheets associated with the shock front.
The radio emission is well correlated with the nonthermal X-rays
\citep{reg86, lor03}. Both show an abrupt turn-on with coincident
edges in the radial directions, but the X-ray emission has more
pronounced narrow peaks. This points to a scenario in which the rims
are limited by the radiative losses\footnote{Note however that such a
model is incapable of reproducing the observed sharp turn-on of radio
synchrotron emission at SN 1006's outer edge, as in Tycho
\citep{cah07}.} in a highly turbulent and amplified magnetic field
\citep{bek02, bek03, elc05, ba06}.  Finally, the observed synchrotron
morphology in SN 1006 is best explained if the bright limbs are polar
caps with the ambient magnetic field parallel to the shock velocity
\citep{rob04}. In the caps, the maximum energy reached by the
accelerated particles, as well as their number, must be higher than
elsewhere \citep{rob04}.

In contrast to the nonthermal X-ray emission, the very faint thermal
X-ray emission seems to be distributed more or less uniformly
\citep{rob04}. This is best seen in the oxygen band (0.5-0.8 keV). It
is, however, difficult to separate the thermal emission from the
nonthermal emission in the caps. Moreover, it is not clear whether the
thermal X-rays should be attributed to the shocked ambient medium
\citep{yak07} or the shocked ejecta \citep{lor03}, but both the
over-solar abundances required to fit the X-ray spectra in the inner
northwest (NW) and NE parts of the SNR and the clumpiness on a
$30\arcsec$ to $1\arcmin$ scale of the low-energy X-ray emission favor
an ejecta origin. The SE region seems to be slightly
different than the rest of the SNR in terms of ejecta composition and
clumpiness too. There is, in particular, evidence for the presence of
cold \citep{wil05} and reverse-shock heated \citep{yak07} iron. In
terms of dynamics, there is evidence for ejecta extending
to/overtaking the BW in the NW \citep{lor03, vil03b_SN1006, rak07}.

In the optical, the remnant has a very different morphology.  Deep
H$\alpha$ imaging reveals very faint Balmer emission around almost the
entire periphery with a clear arc or shock front running from east to
south \citep{wil97b_SN1006, wig03}. Interestingly, no synchrotron
emission is detected in X-ray in this southeast (SE) region whereas
extremely faint (and highly polarized) synchrotron emission is
detected in the radio \citep{reg93}. Besides, none of the optical
emission is of synchrotron origin. There is also a clear Balmer line
filament in the NW of SN 1006 which has been observed in great detail
in the optical \citep{ghw02_SN1006, sog03, rak07}, ultraviolet
\citep{rab95, kor04, haf07} and X-ray bands \citep{lor03,
vil03b_SN1006, acb07}, providing diagnostics for the shock speed ($V_s
\sim 2400-3000\: \mathrm{km/s}$), ion-electron thermal equilibration
at the shock ($T_e/T_p \le 0.07$) and preshock ambient density ($0.25
\: \mathrm{cm}^{-3} \le n_0 \le 0.4 \: \mathrm{cm}^{-3}$). Optical
proper motions \citep{wig03} combined with the shock speed estimate
\citep{ghw02_SN1006, hem07} along this NW filament leads to a distance
to the SNR between $1.8$ and $2.3$ kpc. However, because the remnant
is interacting with a denser medium in the NW \citep{mog93, dug02},
the shock speed and the ambient density, in particular, might be
respectively higher and lower elsewhere \citep{acb07}. For instance,
the broadening of the oxygen \citep{vil03b_SN1006} and silicon
\citep{yak07} X-ray emission-lines suggest a shock speed $V_s \ge 4000
\: \mathrm{km/s}$ (although these measurements were carried out at
different locations in SN 1006). Such high velocities are compatible
with recent proper motion measurements based on radio data
\citep{moc04, rec08}. Moreover, the lack of observed TeV $\gamma$-ray
emission \citep{aha05_SN1006} sets an upper limit on the ambient
density of $n_0 \le 0.1 \: \mathrm{cm}^{-3}$ \citep{ksb05} and the
high latitude ($\sim 500 \: \mathrm{pc}$ at a distance of $2$ kpc) of
SN 1006 suggests that $n_0 \sim 0.03-0.04 \: \mathrm{cm}^{-3}$
\citep{fe01_ISM}.  

\section{Data\label{sect-data}}

\subsection{X-ray\label{subsect-data-xray}}

We used the most recent \textit{Chandra} data of SN 1006 (ObsId 3838
and ObsId 4385 up to 4394) which were obtained in April 2003 with the
ACIS-I imaging spectrometer in timed exposure and very faint data
modes. Eleven pointings were necessary to cover the entire extent of
the remnant. The final image is shown in Figure
\ref{fig_SN1006_3color_xray}. The X-ray analysis was done using CIAO
software (ver. 3.4). Standard data reduction methods were applied for
event filtering, flare rejection, gain correction. The final exposure
time amounts to $\sim 20$ ks per pointing.

\subsection{Radio\label{subsect-data-radio}}

Radio observations were performed in 2003 (nearly at the same
time as the X-ray observations) with both the Australia Telescope
Compact Array (ATCA) and the Very Large Array (VLA).

The ATCA consists of an array of six 22-m antennae that attain a
maximum baseline of 6 km in the East-West direction.  The ATCA
observations occurred on three separate occasions in antenna
configurations that optimize high spatial resolution: January 24 in
its 6-km B configuration for 12 hours, March 3 in its 6-km A
configuration for 12 hours, and June 12 in its 750-m C configuration
for 7 hours. Observations were made using two 128-MHz bands (divided
into 32 channels each) centered at 1384 and 1704 MHz.  The source PKS
1934--638 was used for flux density and bandpass calibration, while PKS
1458--391 was used as a phase calibrator.  The total integration time
on SN 1006 was over 1500 minutes.

The VLA consists of 27 25-m antennae arranged in a ``Y'' pattern.
Observations with the VLA were carried out during 4 hours on January
25th in its hybrid CnD configuration (arranged to maximize spatial
resolution when observing southern declination sources).  Observations
were made using two 12.5-MHz channels centered at 1370 and 1665 MHz.
3C286 and 1451--400 were observed for flux and phase calibration.  The
total VLA integration time on SN 1006 was 140 minutes.

Data from all of the observations were combined, then uniformly
weighted during imaging to minimize the effects of interferometric
sidelobes.  The final image, shown in Figure \ref{fig_SN1006_radio},
has a resolution of $\sim 6\arcsec \times 9\arcsec$ and an
off-source rms noise of $20 \: \mu\mathrm{Jy/beam}$ near the rims.
Since the angular size of SN 1006 is comparable to the primary beam of
both instruments (half a degree), corrections were applied to the
image to recover the lost flux. After primary beam corrections, the
total flux density was recovered to within $< 5\%$ of the expected value. 
With primary beam correction, noise increases in the faint SE portion.
For the study of the azimuthal variations of the radio emission at the shock, 
we use non beam corrected data (\S \ref{subsubsect-azimu-synch-emiss}).

\subsection{Optical\label{subsect-data-optical}}

We used the very deep H$\alpha$ image presented by \cite{wig03}.  This
image taken in June 1998, that is 5 years before the X-ray data, is shown
in Figure \ref{fig_SN1006_Halpha}.
To compare the optical and X-ray images, we had to correct for the
remnant's expansion. In the optical, proper motions of $0.280\pm0.008
\: \arcsec \: \mathrm{yr}^{-1}$ were measured (from April 1987 to June
1998) along the NW rim where thin nonradiative Balmer-dominated
filaments are seen \citep{wig03}. In the radio, an overall expansion
rate of $0.44\pm0.13 \: \arcsec \: \mathrm{yr}^{-1}$ was measured
\citep[from May 1983 to July 1992,][]{mog93}, although higher values
have been measured using more recent observations \citep{moc04,
rec08}. This value does not include the NW rim where the optical
filaments are observed because there is simply little or no radio
emission there.  The observed expansion rate is clearly higher in the
radio than in the optical and is consistent with the picture in which
the BW encounters a localized and relatively dense medium in the NW. 
For simplicity, when using radial profiles, we use the value of
$0.40\arcsec \: \mathrm{yr}^{-1}$, so that the optical profiles are
shifted by $2\arcsec$ to be compared with the X-ray profiles. As we
show below, this correction is negligible compared to other
uncertainties when trying to locate the fluid discontinuities.

\section{Key observational results\label{sect-results}}

\subsection{Fluid discontinuities\label{sub-sect-radii-SE}}

In this section, we determine the location of the fluid
discontinuities -- i.e., the BW and CD -- in the SE quadrant of SN
1006. Our goal is to measure the gap between the BW and CD
as a function of azimuth.

To trace the location of the BW, we use the H$\alpha$ image of SN 1006
which shows very faint filaments of Balmer emission around much of the
periphery of the remnant (Fig.~\ref{fig_SN1006_all}, \textit{top-right
panel}).  In the southeastern quadrant, the H$\alpha$ emission follows
a nearly circular arc.  We extracted radial profiles by azimuthally
summing over $4^{\circ}$ wide sectors starting from $158^{\circ}$
(from west) to $302^{\circ}$ (in the clockwise direction).  The center
of these profiles was determined to provide the best match to the
curvature of the H$\alpha$ rim.  The coordinates of the center are
$(\alpha_{\mathrm{J2000}}, \delta_{\mathrm{J2000}}) =
(15^{\mathrm{h}}02^{\mathrm{m}}56.8^{\mathrm{s}},-41^\circ 56\arcmin
56.6\arcsec)$. This center is very close to the geometrical center of
the SNR, which, given its overall circularity, may not be too far from
the overall expansion center which has not yet been determined. In
practice we found it very difficult to determine the location of the
optical rim based purely on a numerical value (e.g., contour value or
enhancement factor above the local background level) because of the
presence of faint stellar emission, poorly subtracted stars, and faint
diffuse H$\alpha$ emission across the image.  So, we identified
locations where the H$\alpha$ emission increased slightly in the
radial profiles by eye and then further checked (again visually) the
corresponding radii on the optical image. We associated generous
uncertainties with the radii derived from this procedure: $12 \arcsec$
outward and $24 \arcsec$ inward (our process tended to overestimate
the radii, hence the asymmetric errors).

To trace the location of the CD, we use the X-ray image in the
low-energy band ($0.5-0.8$ keV) which contains most of the oxygen
lines (Fig. \ref{fig_SN1006_all}, \textit{bottom-left panel}). While
it is not clear whether the oxygen emission comes from the shocked
ambient medium or the shocked ejecta, we consider here an ejecta
origin. We discuss and justify this assumption below (\S
\ref{subsect-oxygen-ejecta-or-ISM}). Another issue comes from the fact
that the X-ray emission in the oxygen band may contain some nonthermal
contribution in the bright limbs. The three-color composite X-ray
image (Fig. \ref{fig_SN1006_3color_xray}) reveals that the oxygen emission
(red color) is in fact fairly different from the nonthermal X-rays
(white color), and appears notably clumpier even in the synchrotron
limbs (see the eastern limb for instance). This tells us that we can
still use the oxygen emission to trace the CD in the bright limbs, at
least in the azimuthal range we selected ($158^{\circ}-302^{\circ}$)
where the synchrotron emission is not completely overwhelming.  Radial
profiles were extracted from a flux image, summing over $1^{\circ}$
wide sectors, with the same center and azimuthal range as for the
optical. In the radial profiles, we selected the radii for which the
brightness becomes larger than $1.5 \times 10^{-5}$
ph/cm$^2$/s/arcmin$^2$.

Figure \ref{fig_radius_Ha_xray} summarizes the above-described
measurements and shows the azimuthal variations of the radii as
determined from the H$\alpha$ emission (\textit{black lines}) and
low-energy ($0.5-0.8$ keV, \textit{red lines}), mid-energy ($0.8-2.0$
keV, \textit{green lines}) and high-energy ($2.0-4.5$ keV,
\textit{blue lines}) X-ray emissions. The cross-hatched lines
indicate the location of the bright synchrotron limbs. They correspond
to the places where the contour of the outer high-energy X-ray
emission (\textit{blue lines}) can be determined. Between the two
synchrotron caps (angles between $200^\circ$ and $270^\circ$), there
is a clear gap between the BW and CD which is easily seen in Figure
\ref{fig_SN1006_all} (\textit{bottom-left} panel).  
In the bright limbs, however, the BW and CD are
apparently globally coincident.  Fingers of ejecta (indicated by the
\textit{stars}) are even sometimes visible clearly ahead of the
H$\alpha$ emission. In these places, the H$\alpha$ emission might not
be the best tracer of the BW location. The mid-energy X-ray emission
(\textit{green lines}) shows that these fingers are also present.  It
is not clear whether this emission traces the BW or, again, the ejecta
(mostly silicon).  The high-energy X-ray emission (\textit{blue
lines}) cannot be used, nor a X-ray spectral analysis, to answer this
point because of the low number of X-ray counts in this region.

Figure \ref{fig_Rs_o_Rc} shows the ratio of radii for the H$\alpha$
and low-energy X-ray emission, which to first approximation
corresponds to the ratio of radii between the BW and CD, $R_{\rm
BW}/R_{\rm CD}$.  With increasing azimuthal angles the ratio of radii
increases from values near unity (within the northeastern
synchrotron-emitting cap) to a maximum of
$R_{\mathrm{BW}}/R_{\mathrm{CD}} \simeq 1.10^{+0.02}_{-0.04}$ before
falling again to values near unity (in the southwestern
synchrotron-emitting cap). Over the entire azimuthal region where the
synchrotron emission is faint, the azimuthally averaged ratio of radii
is $R_{\rm BW}/R_{\rm CD} \simeq 1.04\pm0.03$. In the regions within
the synchrotron rims, $R_{\rm BW}/R_{\rm CD} \simeq 1.00$.

\subsection{Synchrotron emission at the blast wave\label{subsubsect-azimu-synch-emiss}}

In this section, we measure the azimuthal variations of the
synchrotron flux extracted at the BW and we investigate how these
variations compare when measured at different frequencies.

For that purpose, we use radio and X-ray data (Fig.
\ref{fig_SN1006_all}, \textit{left panels}).  We extracted the flux
in a $30\arcsec$-wide region behind the BW along the SE rim. To define
the BW location, we used the well-defined radii obtained from the
high-energy X-ray image for the northeastern ($\theta \leq 200^\circ$)
and southwestern ($\theta \geq 270^\circ$) regions while, in the
remaining SE quadrant ($200^\circ < \theta < 270^\circ$), we used the
radii determined from the H$\alpha$ image since no synchrotron
emission is detected there.

In Figure \ref{fig_flux_vs_angle_radio_xrays} (\textit{top panel}), we
show the azimuthal variations of the projected brightness in the radio
(1.5 GHz, \textit{black lines}), and several X-ray energy bands:
$0.5-0.8$ keV (\textit{red lines}), $0.8-2.0$ keV (\textit{green
lines}) and $2.0-4.5$ keV (\textit{blue lines}). The radio brightness
was multiplied by $2 \times 10^{-5}$ to make it appear on the same
plot with the X-ray data.  When moving from the northeastern or
southwestern synchrotron-emitting caps toward the faint SE region, we
see that the projected brightness gradually decreases in both radio
and X-ray bands. The radio, medium- (\textit{green lines}) and
high-energy (\textit{blue lines}) X-rays show mostly the variations of
the synchrotron emission, while the low-energy X-rays (\textit{red
lines}) show mostly the variations of the thermal (oxygen) emission,
but may contain some nonthermal contribution as well.

In Figure \ref{fig_flux_vs_angle_radio_xrays} (\textit{bottom panel}),
we show the same azimuthal variations but rescaled roughly to the same
level in the synchrotron-emitting caps. The logarithmic scale shows
that the brightness contrast between the bright caps and faint SE
region is of order $20$ to $50$ in the radio and X-rays. In fact, this
value might even be larger since the radio and high-energy X-ray
fluxes in the faint SE quadrant are compatible with no emission.  We
note that the azimuthal variations of the radio synchrotron emission
are much less pronounced than those in the X-rays. 

\section{Relating hydrodynamical models to the observations\label{sect-models-vs_obs}}

In the previous section, we have found evidence for azimuthal
variations of the separation between the BW and CD and synchrotron
brightness just behind the BW (\S \ref{sect-results}). Both the
separation and the brightness are correlated: the brighter the BW, the
smaller the separation. We measured in particular an unexpectedly small
separation between the BW and CD ($\sim 1.04$) in the SE quadrant of
SN 1006 where little to no synchrotron emission is detected.  These
measurements depend on the assumption that the oxygen emission comes
from the shocked ejecta.  After justifying this approach below (\S
\ref{subsect-oxygen-ejecta-or-ISM}), we first investigate the
different scenarios that could potentially lead to a small ratio
of radii between the BW and CD assuming no CR acceleration at the
shock (\S \ref{subsect-no-CR-accel}). Finally, we interpret the
observed azimuthal variations of the ratio of radii and synchrotron
brightness using CR-modified hydrodynamic models (\S
\ref{subsect-CR-accel}).

\subsection{Does the oxygen come from the ejecta?\label{subsect-oxygen-ejecta-or-ISM}}

A strong assumption made in the measurement of the ratio of radii
between the BW and CD in SN 1006 is that the thermal X-ray emission
(from the oxygen) is associated with the shocked ejecta.  

There is, however, still the possibility that the oxygen emission is
associated with the shocked ambient medium. \cite{yak07} have shown
that the integrated X-ray spectrum of the SE quadrant can be described
by a combination of three thermal plasmas in non-equilibrium
ionization and one power-law component.  One of the thermal
components, assumed to have solar abundances and therefore associated
with the shocked ambient medium, was able to produce most of the
observed low-energy X-rays (in particular the K$\alpha$ lines from O
\textsc{vii}, O \textsc{viii} and Ne \textsc{ix}).  The two other
thermal components, with non-solar abundances, were able to reproduce
most of the K$\alpha$ lines of Si, S and Fe, and were attributed to
the shocked ejecta.

There are several arguments from an examination of the high resolution
\textit{Chandra} images, however, that favor an ejecta origin for the
low-energy X-ray emission.  The three-color composite X-ray image
(Fig.~\ref{fig_SN1006_3color_xray}) shows that the spectral character
of the X-ray emission in the SE quadrant (where there is little to no
synchrotron emission) does not change appreciably as a function of
radius behind the BW, which would be expected if the \cite{yak07}
picture were correct. The lack of virtually any X-ray emission between
the BW and CD can be explained by the expected low density of the
ambient medium \citep[$\sim 0.03-0.04$ cm$^{-3}$][]{fe01_ISM} which
can reduce both the overall intensity and the level of line emission
(due to strong nonequilibrium ionization effects). If the soft X-ray
emission were from the shocked ambient medium its brightness should in
principle rise gradually behind the BW as a result of the projection
of a thick shell onto the line-of-sight \citep{wah05}, but the radial
profile of the low-energy X-ray emission (not shown) in fact turns on
rather quickly at the periphery.

Another argument in favor of an ejecta origin is the observed clumpiness
of the oxygen emission throughout the SE region. There is also the
protuberances seen right at the edge of SN 1006, which are suggestive
of Rayleigh-Taylor hydrodynamical instabilities that are expected at
the CD (see Fig. \ref{fig_SN1006_3color_xray}). Note that the outer edge of
the long, thin filament of X-ray emission in the NW -- which overlaps
the brightest H$\alpha$ emission and therefore is likely to be at least
partly due to shocked ambient medium -- is much smoother than the edge of the
remnant in the SE. Finally, high emission measures of oxygen in the
shocked ejecta at young dynamical SNR ages are compatible with most
thermonuclear explosion models. Essentially, all 1-D and 3-D
deflagration models, and all delayed detonation and pulsating delayed
detonation models have oxygen in the outer layers (but not prompt
detonations or sub-Chandrasekhar explosions) \citep{bab03}.

\subsection{The small gap in the faint SE region\label{subsect-no-CR-accel}}

We have presented evidence in \S \ref{sub-sect-radii-SE} for a small
ratio of radii $R_{\mathrm{BW}}/R_{\mathrm{CD}} \sim 1.04$ between the
BW and CD in the SE quadrant of SN 1006.  This is also the location
where the radio and X-ray nonthermal emission are either very weak or
undetected.  The most straightforward explanation for this lack of
synchrotron emission is an absence of efficient CR acceleration at the
BW here. In the following sections, we first investigate whether pure
1-D hydrodynamical models for SNR evolution (i.e., without CR
acceleration) can reproduce such a small ratio of radii.  We consider
the role of ambient density, ejecta profile, and explosion energy on
the size of the gap between CD and BW.  Then we consider the effects
of 3-D projection on the hydrodynamically unstable CD.

\subsubsection{Predictions from standard hydrodynamical models\label{subsubsect-hydro-simu-no-CRs}}

To understand how a small gap can be produced in a region where there
is apparently no evidence for efficient particle acceleration, we used
standard one dimensional (1-D) spherically symmetric numerical
hydrodynamical simulations that follow the interaction of the ejecta
with the ambient medium. These simulations do not include any CR
component. 

Key inputs to these simulations are the initial density profiles of
the ejecta and ambient medium. In the following, we consider a uniform
density in the ambient medium and two different initial ejecta density
profiles: exponential and power-law. The exponential form has been
shown to be most representative of explosion models for thermonuclear
SNe \citep{dwc98}, but we also present results with a power-law
distribution to quantify the impact of the shape or compactness of the
ejecta profile on $R_{\mathrm{BW}}/R_{\mathrm{CD}}$.

In general, an exponential profile is expected for thermonuclear SNe
because the explosion is driven by the continuous release of energy
from the burning front as it propagates through the star, while the
power-law profile (more compact) is expected for core collapse SNe
because the explosion is driven by a central engine (core bounce), and
the shock loses energy as it propagates through the star
\citep{mam99}. Power law profiles with $n=7$ have been used to
represent 1-D deflagration models \citep[in particular, model W7
from][]{not84}, while exponential profiles are more adequate to
represent delayed detonation models \citep{dwc98}. When comparing
these two analytical profiles, it is worth noting that they transmit
momentum to the ISM in a different manner. The more compact power law
profile is a more efficient piston for ISM acceleration, and will lead
to smaller gaps than the less compact exponential profile
\citep{dwc98, bab03}.

The hydrodynamical simulations provide the radii of the BW and CD at
any time for a given ejecta profile, ambient medium density, and
explosion energy. In Figure \ref{fig_n0_vs_ratio}, we quantify the
impact of these three contributions on the ratio of radii
$R_{\mathrm{BW}}/R_{\mathrm{CD}}$. In particular, we plot the ratio
obtained for a wide range of ambient medium density ($0.001 \:
\mathrm{cm}^{-3} \leq n_0 \leq 30 \: \mathrm{cm}^{-3}$) using 1-D
exponential (EXP) and powerlaw (PL7) ejecta profiles with different
kinetic energies. The cross-hatched domains define the range of ratio
of radii\footnote{In Figure \ref{fig_n0_vs_ratio}, it is important to
keep in mind that the maximum value of the ratio found in Tycho comes
from a presumably efficient particle acceleration shock region while
in SN 1006, it comes from an inefficient particle acceleration
region.} observed in SN 1006 and Tycho
($R_{\mathrm{BW}}/R_{\mathrm{CD}} \leq 1.10$) and range of typical
ambient densities ($0.01 \: \mathrm{cm}^{-3} \leq n_0 \leq 0.06 \:
\mathrm{cm}^{-3}$ in SN 1006; $0.1 \: \mathrm{cm}^{-3} \leq n_0 \leq
0.6 \: \mathrm{cm}^{-3}$ in Tycho) consistent with constraints derived
from the observations \citep{acb07, cah07}.  It is clear that the 1-D
hydrodynamical simulations are unable to reproduce a ratio of radii as
small as the one observed in both SNRs.  We note however that such a
comparison between the observations and the models is not
straightforward since our measurements are likely to be affected by
projection and other effects as we detail next.

\subsubsection{Three dimensional projection effects\label{subsubsect-geom-proj}}

One of the limitations of our numerical simulations is the one
dimensionality.  Hydrodynamical simulations in 2-D or 3-D show that
hydrodynamical (Rayleigh--Taylor) instabilities project pieces of
ejecta ahead of the 1-D CD.  In these simulations, the outermost
pieces of ejecta reach half way of the gap between the 1-D CD and BW
\citep[e.g.,][]{chb92, wac01}. Therefore, because of such protrusions,
the line-of-sight projected CD radius, $\hat{R}_{\mathrm{CD}}$, will
appear larger than the true average CD radius, $R_{\mathrm{CD}}$.  On
the other hand, because the BW is not as highly structured as the CD
interface (it is not subject to Rayleigh--Taylor instabilities), it is
reasonable to assume that the true average BW radius,
$R_{\mathrm{BW}}$, is very close to the projected value,
$\hat{R}_{\mathrm{BW}}$.  It follows that the ratio of projected radii
between the BW and CD, $\hat{R}_{\mathrm{BW}} /
\hat{R}_{\mathrm{CD}}$, will be smaller than the true average ratio,
$R_{\mathrm{BW}} / R_{\mathrm{CD}}$. Quantifying the effects due to
projection is of considerable importance for interpreting the ratio of
radii between the BW and CD in the context of particle
acceleration\footnote{It is not even clear how the 1-D CR-modified
hydrodynamical results on the ratio of radii need to be adjusted given
multi-D Rayleigh--Taylor instability effects. Preliminary studies on
this have been done by \cite{ble01} but this work does not fully
quantify all relevant effects.}.

There are several attempts aimed at estimating the projection
correcting factor, $\xi$, where $\xi$ is defined as
$\hat{R}_{\mathrm{CD}} = (1 + \xi) \: R_{\mathrm{CD}}$. Based on the
projection with ejecta protrusions determined from a power-spectrum
analysis done at the observed CD in Tycho, \cite{wah05} estimated an
amount of bias of $\xi_{\mathrm{proj}} \simeq 6\%$. Based on the
projection of a shell of shocked ejecta with protrusions calculated in
a 2-D hydrodynamical simulations, \cite{dw00} and \cite{wac01} found a
slightly larger value of $\xi_{\mathrm{proj}} \simeq 10\%$. Taking
$\xi=10\%$, the predicted ratio of projected radii,
$\hat{R}_{\mathrm{BW}} / \hat{R}_{\mathrm{CD}} = R_{\mathrm{BW}} /
R_{\mathrm{CD}} \: / \: (1 + \xi)$, becomes $1.18 / (1+0.1) \simeq
1.07$, where $1.18$ is the ratio of radii found at an age of 1000
years using 1-D power-law ejecta profiles assuming a kinetic energy of
the explosion of $10^{51}$ ergs and an ambient medium density of $0.03
\: \mathrm{cm}^{-3}$. Using exponential ejecta profiles, we find
$\hat{R}_{\mathrm{BW}} / \hat{R}_{\mathrm{CD}} =1.14$ because they
produce higher ratio of radii (i.e., $R_{\mathrm{BW}} /
R_{\mathrm{CD}}=1.25$). In either case (power-law or exponential
ejecta profiles), the average ratio (resp. 1.07 or 1.14) is still
larger than the average value of 1.04 measured in the SE quadrant of
SN 1006 (cf. \S \ref{sub-sect-radii-SE}). 

There are several possibilities to explain such discrepancy between
the models and the observations. For instance, the value of
$\xi_{\mathrm{proj}}$ could be larger in 3-D than in 2-D. Indeed,
models show that hydrodynamical instabilities can grow considerably
faster (by $\sim 30\%$) and penetrate further in 3-D than in 2-D
\citep{kaa00}. There can also be a certain amount of inhomogeneity in
the ejecta density distribution. Spectropolarimetric observations of
thermonuclear SNe do show that ejecta are clumpy on large scales
\citep{lel05}. Besides, \cite{wac01} have shown that high density
clump traveling with high speed in the diffuse SN ejecta can reach and
perturb the BW. This requires a high density contrast of a factor 100.
The density contrast is certainly not as high in SN 1006 because
otherwise we would see it in the low-energy X-ray and/or in the
optical image, but may be non-negligible. According to the previous
calculation, the ejecta clumping is required to contribute an
additional $\xi_{\mathrm{clump}} \sim 3-10\%$ factor (so that $\xi =
\xi_{\mathrm{proj}}+\xi_{\mathrm{clump}}$) in order to make the
observed and predicted ratios of radii agree. Finally, although we
could not come up with a reasonable solution, we cannot exclude the
possibility of a complex 3-D geometry in the SE quadrant of SN 1006
where the outermost extent of the shocked ejecta would not be directly
connected to the shock front emission in the hydrodynamical sense
(i.e., the CD and BW would corresponds to different parts of the
remnant seen in projection).

\subsection{Azimuthal variations of the CR ion acceleration\label{subsect-CR-accel}}

In this section, we try to explain both the observed azimuthal
variations of the ratio of radii between the BW and CD (\S
\ref{sub-sect-radii-SE}) and synchrotron flux at the BW (\S
\ref{subsubsect-azimu-synch-emiss}) in SN 1006 in the context of
CR-modified shock hydrodynamics where the efficiency of CR
acceleration is varying along the BW.

In the following, we use CR-modified hydrodynamic models of SNR
evolution which allow us for a given set of acceleration parameters to
calculate the ratio of radii, $R_{\mathrm{BW}}/R_{\mathrm{CD}}$, and
the synchrotron flux, $F_{\nu}$, at any frequency $\nu$. Relevant
parameters are the injection rate of particles into the DSA process,
the magnetic field strength and the particle diffusion coefficient (or
equivalently the turbulence level). Our goal will not consist in
finding the exact variations of these parameters along the shock by
fitting the data, but rather to provide a qualitative description of
the variations based on the theory to verify whether the model
predictions are consistent with the observations or not.

\subsubsection{Conceptual basis for modeling\label{subsect-basis-modelling}}

Here, we explicate the various assumptions concerning the spatial
distribution of the ambient density and magnetic field before running
our CR-modified hydrodynamic models. The knowledge of the remnant's
environment is important since it affects its evolution and also its
emission characteristics.

First, we assume that the ambient density is uniform. This is
suggested by the quasi-circularity of the H$\alpha$ filament in the SE
of SN 1006 and a reasonable assumption for remnants of thermonuclear
explosion \citep{bah07}. Second, we assume that the ambient magnetic
field direction is oriented along a preferred axis. In SN 1006, this
would correspond to southwest-northeast axis which is the direction
parallel to the Galactic plane \citep[following][]{rob04}. The
magnetic field lines are then parallel to the shock velocity in the
brightest part of the synchrotron caps in SN 1006 and perpendicular at
the equator. The angle between the ambient magnetic field lines and
the shock velocity will determine the number of particles injected in
the DSA process, i.e., the injection rate. The smaller the angle, the
larger the injection rate \citep{elb95, vob03}.

With this picture of an axisymmetry around the magnetic field
orientation axis in mind, the azimuthal conditions at the BW are
essentially spatially static, the evolution is self-similar and hence
temporal evolution of gas parcels can be followed with a 1-D
spherically symmetric code. We will run such a code with specific
initial conditions at each azimuthal angle along the BW (viewing the
SNR in a plane containing the revolution axis of the magnetic field).
We implicitly assume a radial flow approximation in our approach,
i.e., that each azimuthal zone is sufficiently isolated from the
others so that they evolve independently.

We run the self-similar models assuming a power-law density profile in
the ejecta with an index of $n=7$, an ejected mass and kinetic energy
of the ejecta of 1.4 $M_{\odot}$ and $10^{51}$ ergs, respectively,
which are standard values for thermonuclear SNe. We assume that the
SNR evolves into an interstellar medium whose  density is $n_0 = 0.03$
cm$^{-3}$ and pressure 2300 K cm$^{-3}$. For such a low density, the
use of the self-similarity is well justified as shown in Figure
\ref{fig_n0_vs_ratio}: the ratio of radii
$R_{\mathrm{BW}}/R_{\mathrm{CD}}$ obtained with a 1-D purely numerical
hydrodynamic simulation with the same initial conditions remains
indeed more or less constant even until the age of 1000 yrs
(\textit{solid yellow lines}). Other input parameters will be
specified later. First, we describe the CR
acceleration model used with the hydrodynamical model.

\subsubsection{CR acceleration model\label{CR-hydro-model}}

The models consist of a self-similar hydrodynamical calculation
coupled with a nonlinear diffusive shock acceleration model, so that
the back-reaction of the accelerated particles at the BW is taken into
account \citep{dee00}. For a given injection rate of protons,
$\eta_{\mathrm{inj}}$, far upstream magnetic field,  $B_{\mathrm{u}}$,
and diffusion coefficient, $D$, the shock acceleration model
determines the shock jump conditions and the particle spectrum from
thermal to relativistic energies \citep{bee99}. This allows us to
compute the ratio of radii and, by following the downstream evolution
of the particle spectrum associated with each fluid element, the
synchrotron emission within the remnant \citep{cad05}.

{The CR proton spectrum at the BW is a piecewise power-law
model with an exponential cutoff at high energies:
\begin{equation} \label{fp}
f_{\mathrm{p}}(E) = a \:  E^{-\Gamma(E)} \:
 \exp \left( -
 E / E_{\mathrm{p,max}} \right),
\end{equation}
where $a$ is the normalization, $\Gamma$ is the power-law index which
depends on the energy $E$, and $E_{\mathrm{p,max}}$ is the maximum
energy reached by the protons. Typically three distinct energy regimes
with different $\Gamma$ values are assumed \citep{bee99}.  The
normalization, $a$, is given by:
\begin{equation}\label{a}
a =  \: \frac{n_{\mathrm{inj}} \: q_{\mathrm{sub}}}{4 \: \pi \: p_{\mathrm{inj}}^3}
\end{equation}
where $n_{\mathrm{inj}} = \eta_{\mathrm{inj}} \: n_0 \
(r_{\mathrm{tot}}/r_{\mathrm{sub}} )$ is the number density of gas
particles injected in the acceleration process, with
$r_{\mathrm{tot}}$ and $r_{\mathrm{sub}}$ being the overall density
and subshock compression ratios, $q_{\mathrm{sub}} = 3 \;
r_{\mathrm{sub}} / (r_{\mathrm{sub}} - 1)$ and $p_{\mathrm{inj}}$ is
the injection momentum. Here, $p_{\mathrm{inj}} = \lambda \:
m_{\mathrm{p}} \: c_{s2}$ where $\lambda$ is a parameter (here fixed
to a value of 4) which encodes all the complex microphysics of the
shock and $c_{s2}$ is the sound speed in the shock-heated gas
immediately downstream from the subshock \citep[see \S 2.2
in][]{bee99}. We note that because there is a nonlinear reaction on
the system due to the injection, the parameter $\eta_{\mathrm{inj}}$
should be in fact related to $p_{\mathrm{inj}}$ \citep[see \S 5
in][for more details]{blg05}.

The CR electron spectrum at the BW is determined by assuming a certain
electron-to-proton density ratio at relativistic energies,
$K_{\mathrm{ep}}$, which is defined as the ratio between the electron
and proton distributions at a regime in energy where the protons are
already relativistic but the electrons have not yet cooled radiatively
\citep[e.g.,][]{elb00}. In the appropriate energy range, the CR
electron spectrum is then:
\begin{equation} \label{fe}
f_{\mathrm{e}}(E) = a \: K_{\mathrm{ep}} \: E^{-\Gamma(E)} \:
 \exp \left( - 
 E / E_{\mathrm{e,max}} \right),
\end{equation}
where $E_{\mathrm{e,max}}$ is the maximum energy reached by the
electrons, which could be eventually lower than that of protons
($E_{p,\mathrm{max}}$) through synchrotron cooling of electrons
depending on the strength of the post-shock magnetic field.
$K_{\mathrm{ep}}$ is left as a free parameter. We will obtain
typically $K_{\mathrm{ep}} \sim 10^{-4}-10^{-2}$.

The maximum energies $E_{p,\mathrm{max}}$ and $E_{e,\mathrm{max}}$
contain information on the limits of the acceleration. They are set by
matching either the acceleration time to the shock age or to the
characteristic time for synchrotron losses, or by matching the
upstream diffusive length to some fraction, $\xi_s$, of the shock
radius, whichever gives the lowest value. We set $\xi_s=0.05$,
a value that allows to mimic the effect of an expanding spherical
shock  as compared to the plane-parallel shock approximation assumed
here \citep[see][]{be96, elb00}. A fundamental parameter in
determining the maximum energy of particles is the diffusion
coefficient, $D$, which contains information on the level of
turbulence and encodes the scattering law \citep{pam06}. For the sake
of simplicity, we assumed the Bohm regime for all particles (i.e.,
diffusion coefficient proportional to energy). We allow deviations
from the Bohm limit (which corresponds to a mean free path of the
charged particles equal to the Larmor radius, which is thought to be
the lowest possible value for isotropic turbulence) via the parameter
$k_0$ defined as the ratio between the diffusion coefficient, $D$, and
its Bohm value, $D_{\mathcal{B}}$. Hence $k_0 \equiv D /
D_{\mathcal{B}}$ is always $\geq 1$ and $k_0=1$ corresponds to the
highest level of turbulence. The smaller the diffusion coefficient (or
$k_0$), the higher the maximum energy (the maximum energy scales as
$1/\sqrt{k_0}$ in the radiative loss case and as $1/k_0$ in the age-
and escape-limited cases).

\subsubsection{Heuristic models}

In order to make predictions for the azimuthal variations of the ratio
of radii and synchrotron flux, we must make some assumptions about the
azimuthal variations of the input parameters used in our CR-modified
hydrodynamical models. Relevant input parameters are the injection
rate ($\eta_{\mathrm{inj}}$), the upstream magnetic field
($B_{\mathrm{u}}$), the electron-to-proton ratio at relativistic
energies ($K_{\mathrm{ep}}$) and the level of turbulence or diffusion
coefficient relative to the Bohm limit ($k_0$).

To understand the influence of these four basic input parameters, we
begin first with a heuristic discussion to set the stage for later
detailed and more physical models (\S \ref{subsubsect-good-model}).
In Figure \ref{fig_heuristic_model_profiles}, we show three models
where we vary the two most important parameters, $\eta_{\mathrm{inj}}$
and $B_{\mathrm{u}}$, with azimuthal angle ($K_{\mathrm{ep}}$ is held
constant and $k_0$ is set to 1 at all angles). Model 1 (\textit{left
panels}) has $\eta_{\mathrm{inj}}$ varying from $10^{-5}$ to
$10^{-3}$, while $B_{\mathrm{u}}$ is held constant at the value $25 \;
\mu\mathrm{G}$. Model 2 (\textit{middle panels}) has constant
$\eta_{\mathrm{inj}} = 10^{-4}$ and $B_{\mathrm{u}}$ varying from $3
\; \mu\mathrm{G}$ to $25 \;\mu\mathrm{G}$. Model 3  (\textit{right
panels}) is a combination of models 1 and 2 and has
$\eta_{\mathrm{inj}}$ varying from $10^{-5}$ to $10^{-3}$ and
$B_{\mathrm{u}}$ varying from $3 \; \mu\mathrm{G}$ to $25
\;\mu\mathrm{G}$.

Note that in all these models, the functional form of the profiles
when $\eta_{\mathrm{inj}}$ and $B_{\mathrm{u}}$ vary is rather
arbitrary and just used for illustration purposes (for completeness we
note that the angular range shown corresponds to the data extraction
region in Fig.~\ref{fig_flux_vs_angle_radio_xrays}). Moreover, the
upstream magnetic field, $B_{\mathrm{u}}$ (\textit{second panels,
solid lines}), is allowed to vary with azimuthal angle because we
implicitly assume that it can be significantly amplified by the
CR-streaming instability. Our model does not include self-consistently
the magnetic field amplification believed to occur at SNR shocks, but
is provided with a simple compression.  Assuming that the magnetic
turbulence is isotropic ahead of the shock, the immediate post-shock
magnetic field, $B_{\mathrm{d}}$ (\textit{dashed lines}), will then be
larger than upstream by a factor $r_B \equiv B_{\mathrm{d}} /
B_{\mathrm{u}} = \sqrt{(1 + 2 \: r_{\mathrm{tot}}^{2})/3}$. This
relation is assumed at each azimuthal angle\footnote{Here, the
compression does not depend on the angle between the upstream magnetic
field and shock velocity as considered in \cite{re98} or \cite{orb07}.
Instead we assume that after the amplification process, the magnetic
field becomes entirely turbulent, so its originally ordered character
has been lost.}.

We show the predicted profiles of the ratio of radii between the BW
and CD (\textit{third panels}) and synchrotron brightness projected
along the line-of-sight in the radio and different X-ray energy bands
(\textit{bottom panels}). We explain later how those brightness
profiles were precisely constructed (\S \ref {subsubsect-good-model}).
Clearly, model 2 does not predict the expected correlation between the
ratio of radii and the synchrotron brightness and hence can be
rejected immediately\footnote{Looking at the BW to CD ratio in
model 2, it naively seems surprising to have a less modified shock
(i.e., higher BW to CD ratio) where the magnetic field is larger.
This effect is due to the presence of Alfv\'en heating in the
precursor \cite[see][]{bee99}, although there is debate about whether
this is the most efficient mechanism for turbulent heating in the
precursor \citep[see][]{amb06}.}. On the other hand, models 1 and 3
lead to the expected correlation: the smaller the BW/CD ratio, the
brighter the synchrotron emission. Comparing models 1 or 3 with model
2 allows us to see that only an azimuthal variation of the injection
rate leads to the appropriate variation in the ratio of BW to CD radii
(\textit{third panels}). Comparing model 1 with model 3 shows the
impact of the magnetic field variations through synchrotron losses on
the azimuthal profiles of the X-ray synchrotron emission
(\textit{bottom panels}).

Although models 1 and 3 seem to describe the key observational
constraints fairly well, they suffer weaknesses as regards their
underlying astrophysical premises.  In model 1, it is difficult to
justify why the magnetic field should be enhanced in the region of
inefficient particle acceleration ($\eta_{\mathrm{inj}} \leq 6 \times
10^{-5}$).  The model proposes a post-shock magnetic field there of
$\sim 80 \: \mu\mathrm{G}$, when a value of $10-15 \: \mu\mathrm{G}$ would
more likely reflect the value of the compressed ambient field.  In
model 3, it is difficult to understand why the turbulence level would
saturate (i.e., $k_0 = 1$) everywhere. Hence, we are led to develop
more elaborated models where $k_0$ and/or $K_{\mathrm{ep}}$ are
allowed to vary with azimuthal angle. A variation of either $k_0$ or
$K_{\mathrm{ep}}$ will not significantly affect the ratio of BW to CD
radii, but will only modify the profiles of the synchrotron emission.
In fact, a model with varying $K_{\mathrm{ep}}$ can be already
excluded because this quantity changes the radio and X-ray synchrotron
intensities in the same way while the observed radio and X-ray
profiles (Fig.~\ref{fig_flux_vs_angle_radio_xrays}) vary with azimuth
in different ways. Varying $k_0$ (as we show in \S \ref
{subsubsect-good-model}) does allow us to modify the relative
azimuthal profiles.

\subsubsection{Can CR-modified models provide $R_{\mathrm{BW}}/R_{\mathrm{CD}}=1.00$?\label{ratio-unity}}

Although CR-modified hydrodynamical simulations predict a ratio of
radii $R_{\mathrm{BW}}/R_{\mathrm{CD}}$ smaller than standard
hydrodynamical simulations, they still predict a lower limit on that
ratio that is strictly larger than unity. For instance, Figure
\ref{fig_heuristic_model_profiles} (\textit{third-right panel}) shows
that the modeled ratio of radii where the synchrotron emission is
strong is $\sim 1.13$. This value was obtained assuming an injection
rate of $10^{-3}$ and an ambient magnetic field of 25 $\mu$G. Keeping
all the same inputs, but reducing the magnetic field value to 3
$\mu$G, we obtain the lowest possible value of
$R_{\mathrm{BW}}/R_{\mathrm{CD}}=1.06$.

On the other hand, the observations of SN 1006 show that the ratio of
radii is of order 1.00 in the synchrotron-emitting caps, a value that
is not possible to obtain from our CR-modified model under the DSA
framework.  However, as already mentioned before in our discussions of
the small gap between the BW and CD in the region of inefficient
acceleration (\S \ref{subsect-no-CR-accel}), we need to consider
projection effects due to hydrodynamical instabilities at the CD which
can reduce the gap by $6-10\%$.  This does not fully account for the
difference with the model and, therefore, we may need to invoke
another effect, such as clumping of the ejecta, to fully explain the
offset between the observation and models.  It is important to note
that in both the regions of efficient and inefficient CR acceleration,
we require essentially the same numerical factor to bring the modeled
ratio of radii (for the appropriate model in each case) into agreement
with the observed ratio. In the following, we introduce an ad-hoc
rescaling of the ratio of radii by decreasing the modeled values by a
constant factor ($\sim 13\%$) at all azimuthal angles. By doing this,
we implicitly assume that this factor contains the total contribution
of effects that are unrelated to the CR acceleration process. The
numerical value used depends, of course, on the particular SN
explosion model used. For example, the factor would need to be
increased slightly in the case of an exponential ejecta profile (cf.,
Fig.~\ref{fig_n0_vs_ratio}).

\subsubsection{Toward a good astrophysical model\label{subsubsect-good-model}}

In Figure \ref{fig_model_k0_var_profiles}, we present in more detail
what  we believe is a reasonable astrophysical model. This model
assumes that $\eta_{\mathrm{inj}}$ increases from a value of $5 \times
10^{-5}$ to a value of $10^{-3}$ while at the same time
$B_{\mathrm{u}}$ increases from a value of $3 \;\mu\mathrm{G}$ to a
value of $20 \; \mu\mathrm{G}$ and $k_0$ increases from an arbitrary
value of 100 to 1 ($K_{\mathrm{ep}}$ is kept constant at all angles).
The azimuthal variations of these parameters is consistent with the
picture of DSA where, once the magnetic field fluctuations at the
shock exceed the background ISM fluctuations, any low injection rate
leads to growth of the turbulent magnetic field in the upstream
region, which in turns leads to an increase of the injection rate
\citep{vob03}. The acceleration efficiency, defined as the
fraction of total energy flux crossing the shock that goes into
relativistic particles \citep[see \S 3.3 in][]{bee99}, is $51\%$ in
the case with $\eta_{\mathrm{inj}} = 10^{-3}$ and $7\%$ in the case
with $\eta_{\mathrm{inj}} = 5 \times 10^{-5}$.

Note that the form of the azimuthal profiles of $\eta_{\mathrm{inj}}$
and $B_{\mathrm{u}}$ is again arbitrary (here a squared-sine
variation) and just used for illustration purpose.  Our goal is not to
find the exact form of the profiles based on an accurate fit of the
model predictions to the observations, but rather to describe
qualitatively how the acceleration parameters vary along the BW in SN
1006. The extremum values for $\eta_{\mathrm{inj}}$ were chosen so
that the predicted variations of the ratio of radii
$R_{\mathrm{BW}}/R_{\mathrm{CD}}$ roughly matches the observations. A
maximum (minimum) value of $\eta_{\mathrm{inj}}$ between $10^{-4}$ and
$10^{-2}$  (resp. $10^{-5}$ and $5 \times 10^{-5}$) will not
significantly affect our results concerning the profile of
$R_{\mathrm{BW}}/R_{\mathrm{CD}}$, but differences will be seen in the
absolute intensity in the synchrotron emission. The maximum value of
$B_{\mathrm{u}}$ was adjusted so that it yields an immediate postshock
value, $B_{\mathrm{d}} \simeq 80 \: \mu{\mathrm{G}}$ (\textit{panels
e}), consistent with the value derived from the thickness of the X-ray
synchrotron-emitting rims in SN 1006, under the assumption that this
thickness is limited by the synchrotron losses of the highest energy
electrons \citep[e.g.,][]{ba06}. Finally, we adjusted the minimum
value of $B_{\mathrm{u}}$ in the faint region so that the predicted
and observed radio synchrotron flux are roughly the same; here we show
profiles with a value of $B_{\mathrm{u}} = 3 \: \mu{\mathrm{G}}$,
i.e.,  the typical ISM value.

In Figure~\ref{fig_model_k0_var_profiles} (\textit{panel g}), we show
the predicted azimuthal profile of ratio of radii between the BW and
CD (\textit{dashed lines}). The ratio of radii was normalized
(\textit{solid lines}) so that it roughly equals 1.00 in the region of
efficient CR acceleration as in the observations. This normalization
procedure (which we justified in \S \ref{ratio-unity}) works
remarkably well. The variations of the ratio of radii reflects nothing
but the variations of the overall density compression ratio,
$r_{\mathrm{tot}}$ (\textit{panel d}). For low injection rates
($\eta_{\mathrm{inj}} \leq 6 \times 10^{-5}$), the compression ratio
comes close to the value obtained in the test-particle case (i.e.,
$r_{\mathrm{tot}} = 4$) and as the injection rate increases, the
plasma becomes more compressible with $r_{\mathrm{tot}} \sim 6$.
However, once the magnetic field has been considerably amplified
($B_{\mathrm{u}} \geq 10 \; \mu\mathrm{G}$), $r_{\mathrm{tot}}$ starts
to slightly decrease due to the heating of the gas by the Alfv\'en
waves in the precursor region \citep[see][]{bee99}.

In Figure~\ref{fig_model_k0_var_profiles} (\textit{panel h}), we show
the azimuthal profiles of the synchrotron emission in the radio and
several X-ray energy bands. To obtain these profiles, we first
computed the radial profile of the synchrotron flux (calculated in a
given energy band) at each azimuthal angle. Then, we projected the
flux along the line-of-sight and finally extracted the projected flux
in $30\arcsec$-wide regions behind the BW (assuming a distance of 2
kpc) as we did in the observations of SN 1006 (\S
\ref{subsubsect-azimu-synch-emiss}). In this procedure, we assumed a
spherically symmetric distribution for the emissivity.  This is a
reasonable assumption as long as we restrict the analysis close to the
BW, considering that the bright limbs in SN 1006 are in fact polar
caps, i.e., where the emissivity distribution is axisymmetric.
Finally, we fixed the radio flux in the synchrotron caps to be the
same as in the observations by adjusting the electron-to-proton
density ratio at relativistic energies, $K_{\mathrm{ep}}$.  This is
obtained for a value of $K_{\mathrm{ep}} = 7 \times 10^{-4}$.

Relaxing the assumption of the Bohm diffusion (i.e., $k_0$) at each
azimuthal angle significantly modifies the azimuthal profiles of the
synchrotron flux in the X-ray band  (while the radio profile is
unchanged) compared to the model~3 of Figure
\ref{fig_heuristic_model_profiles} in which $k_0$ was always equal to
1.  This is because an increase of $k_0$ decreases the maximum energy
of electrons, $E_{e,\mathrm{max}}$. Since the synchrotron emission is
very sensitive to the position of the high-energy cutoff in the
electron distribution (i.e., $E_{e,\mathrm{max}}$), all profiles do
not show anymore a plateau in the region of efficient CR acceleration
but are now gradually decreasing (\textit{panel h}). Besides, the
synchrotron brightness in the X-ray bands now decreases faster than in
the radio (\textit{panel i}). This is roughly consistent with what is
observed in SN 1006. We note that in this model, the maximum energy of
both protons and electrons (\textit{panel f}) decrease rather quickly
as we move toward regions of inefficient CR acceleration as opposed to
the previous model~3 with $k_0=1$ where the maximum energy of
electrons remained approximately constant (not shown). Overall, our
results from this model are consistent with those of \cite{rob04}
which were based on a different approach (i.e., measure of the
azimuthal variations of the cutoff-frequency all around SN 1006) using
\textit{XMM-Newton} observations.

\subsubsection{Other azimuthal profiles for $\eta_{\mathrm{inj}}$, $B_{\mathrm{u}}$ and $k_0$}

We have found that a simple model in which the injection rate of
particles in the acceleration process ($\eta_{\mathrm{inj}}$),
amplified magnetic field ($B_{\mathrm{u}}$) and level of turbulence
($k_0$) were gradually increasing around the shock front was able to
provide predictions in a good agreement with the observations for the
azimuthal variations of the ratio of radii and radio and X-ray
synchrotron fluxes. The input parameters and their azimuthal
variations were chosen based on the physical picture where the
injection of particles and amplification of the magnetic turbulence
are coupled.

One can imagine however more complicated scenario and hence more
complicated input profiles for the acceleration parameters
($\eta_{\mathrm{inj}}$, $B_{\mathrm{u}}$, $k_0$). For instance, the
coupling between the injection and the turbulence which occurs at the
beginning of the acceleration process may end at some point when the
amplitude of the magnetic field fluctuations become so high that their
further growth is prevented by strong dissipation processes
\citep{vob03}. The injection, magnetic field and level of turbulence
may in fact saturate (over some azimuthal range) as we reach regions of
very efficient CR acceleration.  Another example of scenario can be
imagined if the shock velocity is not uniformly distributed along the
shock (as opposed to what we have assumed in our models) due to a
asymmetrical explosion. Such a scenario could explain why the bright
synchrotron-emitting caps in SN 1006 seem to have larger radii than
the faint regions. If so, the injection could be larger in the region
of higher velocity while the level of turbulence and magnetic field
could potentially saturate there. In general, these new scenarios will
not lead to a significant modification of the profile of the ratio of
radii. However, we do expect some changes in the azimuthal profiles of
the radio and X-ray synchrotron flux. They will generally be flatter
in the regions of very efficient CR acceleration.

\section{Discussion\label{sect-discussion}}

\subsection{Polar cap morphology\label{sub-morphology}}

The model in which the injection rate of particles, magnetic field and
level of turbulence are all varying as a function of azimuth was
successful in reproducing the overall azimuthal variations of the
radio and X-ray synchrotron flux observed along the BW of SN 1006
(Fig.~\ref{fig_model_k0_var_profiles}). In fact, it is actually
possible to build a 2-D projected map of the radio and X-ray
synchrotron morphology comparable to the radio and X-ray images of SN
1006 (Figs. \ref{fig_SN1006_3color_xray} and \ref{fig_SN1006_radio}).
This requires us to know the three-dimensional distribution of the
emissivity. Here, we consider polar caps as suggested by \cite{rob04}.

To simplify the calculation, we consider that the radial profiles of
the emissivity obtained from the different values of injection,
magnetic field and turbulence level have an exponential form
characterized by a maximum emissivity at the shock and width $\Delta
R$ over which the emissivity decreases (see Appendix
\ref{app-projection}). We also fixed the BW radius, 
$R_{\mathrm{BW}}$, to be the same at each azimuthal angle where the
injection and magnetic field vary, although it slightly depends on the
CR acceleration efficiency. These are good approximations.

In the case of the radio synchrotron emission,  the width $\Delta R$
does not depend on the azimuthal angle (from the pole to the equator),
and we have typically $\Delta \tilde{R} \equiv \Delta R /
R_{\mathrm{BW}} = 0.01$. The projected radio morphology is shown in
Figure \ref{fig_projection} (\textit{top-right panel}). It has a
bipolar limb-brightened morphology as is observed in SN 1006. 

In the case of the X-ray synchrotron emission, the width $\Delta R$ is
expected first to increase (starting from the equator) as particles
reach higher energies and the emission builds up, and then to slightly
decrease when the effect of radiative losses starts to become
important due to the larger post-shock magnetic field as we move
toward the pole. Because the brightness is much lower at the equator
than at the pole, a reasonable approximation is to consider the width
constant with typically $\Delta \tilde{R} = 2 - 5 \times 10^{-3}$
(\textit{bottom-left panel}) or even a slightly increasing width from
the pole to the equator (\textit{bottom-right panel}). A slightly
increasing width will not change the overall morphology. It still
produces a thin rim whose projected width is decreasing starting from
the pole to the equator. In Figure \ref{fig_projection}, we have
assumed that the emissivity contrast at the BW between the pole and
the equator is always a factor 100. In fact, because the contrast
between the pole and equator is larger in the X-rays than in the
radio, we will obtain a similar morphology but with the size of the
caps slightly reduced in X-ray. In other words, the X-ray synchrotron
emitting rims become geometrically thinner and of smallest azimuthal
extent as we go to higher energy. All these effects would be however
slightly attenuated if we had included the instrumental effects
(particularly the PSF), but overall this is consistent with what is
observed in SN 1006.

\subsection{CR acceleration in a partially ionized medium\label{sect-neutral-med}}

When trying to delineate the BW, we found that the X-ray
synchrotron-emitting rims and the diffuse filament of H$\alpha$
emission were coincident over some little azimuthal range, notably for
instance at the edge of the eastern cap where the synchrotron emission
starts to turn on (see Fig. \ref{fig_SN1006_3color_xray}). SN 1006 is
one of the rare remnants where this characteristic is clearly
observed, although this likely happens in the Tycho SNR too.

In general, H$\alpha$ emission from non-radiative shocks is believed
to arise when the blast wave encounters partially ionized gas. This
points to the presence of neutral atoms in the ambient medium around
SN 1006 \citep{ghw02_SN1006}. Theoretical studies have shown that in
such medium, the scattering Alfv\'en waves should be damped,
henceforth quenching the acceleration of high energy particles
\citep{drd96}. In fact, recent theoretical studies suggest that the
magnetic field could be turbulently amplified even in such medium
\citep{byt05}. The fact that we observe the H$\alpha$ and synchrotron
X-rays to be coincident gives support to this statement. Other
observations of H$\alpha$ emission coexisting with synchrotron X-rays
would be very interesting in that regard.

\subsection{Comparison between SN 1006 and Tycho}

Like SN 1006, the Tycho SNR is believed to be the remnant of a
thermonuclear explosion and, as such, is expected to develop an
approximately exponential ejecta density profile and to evolve in a
uniform ambient medium. Measurements of the gap between BW and CD have
been done in the Tycho SNR using \textit{Chandra} observations
\citep{wah05}. We comment on the Tycho results in the light of our
results.

Because the thin and bright X-ray synchrotron-emitting rims
concentrated at the BW are observed all around Tycho, in contrast to
SN 1006, the BW's location can be easily determined. In addition,
because the X-ray emission from the shocked ejecta is dominant in
Tycho, in contrast to SN 1006, it is also fairly easy to determine the
CD's location. The observed azimuthally-averaged
ratio of radii between the BW and CD was found to be $\sim 1.04$ in
Tycho \citep{wah05}. After correcting, approximately, for effects due
to projection of the highly structured CD, the ratio of radii became
$\sim 1.075$.  This is still a lot smaller than the (unprojected)
ratio of $\sim 1.25$ derived using standard one-dimensional
hydrodynamical simulations with no CR acceleration, at the current age
of Tycho ($430$ yrs).  It was then concluded that efficient CR ion
acceleration was occurring around nearly the entire BW of Tycho.

More recently, \citet{cah07} applied a CR-modified self-similar
hydrodynamic model to the radio and X-ray observations of the Tycho
SNR using the observed ratio of BW to CD radii as an important input
constraint.  Using reasonable values for the hydrodynamical parameters
(ambient medium density, SN explosion energy, ejected mass), DSA
parameters (injection efficiency, magnetic field, diffusion
coefficient, electron-to-proton ratio at relativistic energies), and
the distance, a good description of the observational properties at
the blast wave of Tycho was obtained. This detailed study provides a
self-consistent/coherent picture of efficient CR ion acceleration.

Regarding SN 1006, we have some good news and some bad (compared to
the case of Tycho, where CR-modified models seem to describe the data
well).  The good news is that there is a significant variation in the
ratio of BW to CD radii as a function of azimuthal angle that is
correlated with the varying intensity of the X-ray and radio
synchrotron emission.  Where the gap is large the synchrotron emission
is faint (or not detected) and where the gap is small the emission is
bright.  This is consistent with an azimuthal variation in the effects
of CR-modification to the remnant dynamics.  The bad news is that in
regions of presumably efficient acceleration (i.e., the bright rims),
the ratio of BW to CD radii approaches small values, very close to 1.
The puzzle is that our CR-modified hydrodynamical models are unable to
produce ratio values less than 1.06, while the specific scenarios
presented above produce ratio values closer to $1.12$. In order
to make progress in our study we decided to introduce an ad-hoc
rescaling of the ratio of radii by decreasing the modeled values at
all angles by a constant factor of $\sim 12\%$.  This procedure
works remarkably well and suggests to us that something else,
unrelated to the CR acceleration process (such as 3-D projection
effects and ejecta clumping), may be responsible for the smaller than
expected gap between the BW and CD in SN 1006.

\section{Conclusion\label{sect-conclusion}}

Using a combination of radio, optical (H$\alpha$) and X-ray images, we
have located the positions of the BW and CD radii along the
southeastern sector of SN 1006. We found that with increasing
azimuthal angles the ratio of radii between the BW and CD increases
from values near unity (within the northeastern synchrotron-emitting
cap) to a maximum of about 1.10 before falling again to values near
unity (in the southwestern synchrotron-emitting cap). These variations
reflect changes in the compressibility of the plasma attributed to
variations in the efficiency of the BW at accelerating CR ions and
give strong support to the overall picture that SNR shocks produce
some fraction of Galactic CRs.  However, at the present time we do not
have a detailed explanation for the apparent overall smallness of the
measured ratios of radii in SN 1006: the minimum value predicted by
our CR-modified self-similar dynamical models is 1.06.  In this study
we simply rescaled the ratio of radii by a constant factor
(independent of azimuthal angle) of $\sim 12\%$ with the expectation
that some process, other than CR acceleration itself, was responsible
for driving the edge of the ejecta closer to the BW all along the rim
of SN 1006. The lack of a definitive astrophysical explanation for
this discrepancy is a significant factor limiting our ability
to understand the CR acceleration process in SN 1006. Further research
into the effects of hydrodynamical instabilities at the CD and ejecta
clumping (two possible explanations for the small BW/CD ratio) would
incidentally provide new insights into the acceleration process.

In addition to the azimuthal variations of the ratio of radii between
the BW and CD we also interpreted the variations of the synchrotron
flux (at various frequencies) at the BW using CR-modified hydrodynamic
models.  We assumed different azimuthal profiles for the injection
rate of particles in the acceleration process, magnetic field and
turbulence level.  We found the observations to be consistent with a
model in which these quantities are all azimuthally varying, being the
largest in the brightest regions. Overall this is consistent with the
picture of diffusive shock acceleration. In terms of morphology, we
found that our model was generally consistent with the observed
properties of SN 1006, i.e., a bright and geometrically-thin
synchrotron-emitting rim at the poles and very faint synchrotron
emission at the equator and in the interior.  In the model, the X-ray
synchrotron-emitting rims are geometrically thinner and of smallest
azimuthal extent than the radio rims, which is in broad agreement with
observations.  This is because the most energetic electrons
accelerated at the blast wave lose energy efficiently in the amplified
post-shock magnetic field.  Based on this picture, it would be worth
trying to measure the azimuthal variation of the magnetic field
strength (from the X-ray rim widths) and thereby gain further insight
into the amplification process.

\acknowledgments It's G.C.-C.'s pleasure to acknowledge
\mbox{J.~Ballet}, \mbox{A.~Decourchelle} and \mbox{D.~C.\ Ellison} for
previous enlightening discussions and work. G.C.-C.\ would like also
to thank \mbox{F.~Winkler} for providing the precious H$\alpha$ image.
 C.B.\ thanks John Blondin for making the VH-1 code available (this
code was used to produce Figure \ref{fig_n0_vs_ratio}). E.M.R.\ is
member of the Carrera del Investigador Cient\'ifico of CONICET,
Argentina. E.M.R.\ was supported by grants PIP-CONICET 6433,
ANPCyT-14018 and UBACYT A055 (Argentina). C.B.\ was supported by NASA
through Chandra Postdoctoral Fellowship Award Number PF6-70046 issued
by the Chandra X-ray Observatory Center, which is operated by the
Smithsonian Astrophysical Observatory for and on behalf of NASA under
contract NAS8-03060. Financial support was also provided by Chandra
grants GO3-4066X and GO7-8071X to Rutgers, The State University
of New Jersey.


\appendix

\section{Projection along the line-of-sight in the case of polar caps\label{app-projection}}

Let $r$ be the distance to the center $O$ of a sphere of radius $R_s$,
$\phi$ the azimuthal angle in the $(O\vec{x},O\vec{z})$ plane with $0
\leq \phi \leq 2 \pi$, and $\theta$ the latitude with $- \pi / 2 \leq
\theta \leq \pi / 2$.  The projection of a
spherical emissivity distribution $\mathcal{E}(r,\theta,\phi)$ onto
the plane $(O\vec{x},O\vec{y})$ results in a two-dimensional
brightness distribution:
\begin{equation}\label{brightness}
\mathcal{B}(x,y) = 2 \: \int_{0}^{\ell} \mathcal{E}(r,\theta,\phi) \: dz,
\end{equation}
where $\ell^2 = x_s^2 - x^2$, and $0 \leq x \leq x_s \equiv R_s \:
\cos \theta$ and $0 \leq y \leq R_s$.  At any point of coordinate
$(x,y,z)$, the emissivity is $\mathcal{E}(x,y,z) =
\mathcal{E}(r,\theta,\phi)$ where $r^2 = x^2 + y^2 + z^2$ and $\theta
= \arcsin \left( y / r \right)$ and $\phi = \arccos( x / \sqrt{x^2 +
z^2} )$.  When the emissivity has a symmetry of revolution around the
$O\vec{y}$ axis (e.g., polar caps), there is no dependency on $\phi$.

Let us consider a sphere where the emissivity $\mathcal{E}$ is
radially decreasing from the maximum $\mathcal{E}_{\mathrm{max}}$ at
the surface of the sphere with a characteristic width $\Delta R$. In
the case of an exponential decrease and cylindrical symmetry, we have:
\begin{equation}\label{emissivity}
  \mathcal{E}(r,\theta,\phi) = \mathcal{E}_{\mathrm{max}}(\theta) \:
  \exp \left( \frac{r - R_s}{\Delta R(\theta)} \right).
\end{equation}
In the case of polar caps, $\mathcal{E}_{\mathrm{max}}$ will
be larger at the poles ($\theta=\pi / 2$ and $\theta = - \pi / 2$).

In Figure \ref{fig_projection}, we show the projected morphology
obtained with the emissivity spatial distribution given by Eq.
(\ref{emissivity}) in the range $0 \leq \theta \leq \pi / 2$ assuming
different values and angular dependencies for the width $\Delta
\tilde{R} \equiv \Delta R / R_s$. In those plots, a squared-sine
variation was assumed for ${\log}_{10} \left(
\mathcal{E_{\mathrm{max}}} \right)$, having the value $0.0$ at the
pole and $-2.0$ at the equator. Decreasing the value at the equator
produces similar rims but with a lower azimuthal extent (i.e.,
smaller polar caps).


\newpage



\bibliographystyle{aa} 

\bibliography{ms} 

\begin{thebibliography}{74}
\expandafter\ifx\csname natexlab\endcsname\relax\def\natexlab#1{#1}\fi

\bibitem[{{Acero} {et~al.}(2007){Acero}, {Ballet}, \& {Decourchelle}}]{acb07}
{Acero}, F., {Ballet}, J., \& {Decourchelle}, A. 2007, \aap, 475, 883

\bibitem[{{Aharonian} {et~al.}(2005){Aharonian}, {Akhperjanian}, {Aye},
  {Bazer-Bachi}, {Beilicke}, {Benbow}, {Berge}, {Berghaus}, {Bernl{\"o}hr},
  {Boisson}, {Bolz}, {Borgmeier}, {Breitling}, {Brown}, {Bussons Gordo},
  {Chadwick}, {Chounet}, {Cornils}, {Costamante}, {Degrange},
  {Djannati-Ata{\"\i}}, {O'C.~Drury}, {Dubus}, {Ergin}, {Espigat}, {Feinstein},
  {Fleury}, {Fontaine}, {Funk}, {Gallant}, {Giebels}, {Gillessen}, {Goret},
  {Hadjichristidis}, {Hauser}, {Heinzelmann}, {Henri}, {Hermann}, {Hinton},
  {Hofmann}, {Holleran}, {Horns}, {de Jager}, {Jung}, {Kh{\'e}lifi}, {Komin},
  {Konopelko}, {Latham}, {Le Gallou}, {Lemi{\`e}re}, {Lemoine}, {Leroy},
  {Lohse}, {Marcowith}, {Masterson}, {McComb}, {de Naurois}, {Nolan},
  {Noutsos}, {Orford}, {Osborne}, {Ouchrif}, {Panter}, {Pelletier}, {Pita},
  {P{\"u}hlhofer}, {Punch}, {Raubenheimer}, {Raue}, {Raux}, {Rayner},
  {Redondo}, {Reimer}, {Reimer}, {Ripken}, {Rob}, {Rolland}, {Rowell},
  {Sahakian}, {Saug{\'e}}, {Schlenker}, {Schlickeiser}, {Schuster}, {Schwanke},
  {Siewert}, {Sol}, {Steenkamp}, {Stegmann}, {Tavernet}, {Th{\'e}oret},
  {Tluczykont}, {van der Walt}, {Vasileiadis}, {Vincent}, {Visser}, {V{\"o}lk},
  \& {Wagner}}]{aha05_SN1006}
{Aharonian}, F., {Akhperjanian}, A.~G., {Aye}, K.-M., {et~al.} 2005, \aap, 437,
  135

\bibitem[{{Aharonian} {et~al.}(2007{\natexlab{a}}){Aharonian}, {Akhperjanian},
  {Bazer-Bachi}, {Beilicke}, {Benbow}, {Berge}, {Bernl{\"o}hr}, {Boisson},
  {Bolz}, {Borrel}, {Braun}, {Brion}, {Brown}, {B{\"u}hler}, {B{\"u}sching},
  {Carrigan}, {Chadwick}, {Chounet}, {Coignet}, {Cornils}, {Costamante},
  {Degrange}, {Dickinson}, {Djannati-Ata{\"\i}}, {O'C.~Drury}, {Dubus},
  {Egberts}, {Emmanoulopoulos}, {Espigat}, {Feinstein}, {Ferrero}, {Fiasson},
  {Fontaine}, {Funk}, {Funk}, {F{\"u}{\ss}ling}, {Gallant}, {Giebels},
  {Glicenstein}, {Gl{\"u}ck}, {Goret}, {Hadjichristidis}, {Hauser}, {Hauser},
  {Heinzelmann}, {Henri}, {Hermann}, {Hinton}, {Hoffmann}, {Hofmann},
  {Holleran}, {Hoppe}, {Horns}, {Jacholkowska}, {de Jager}, {Kendziorra},
  {Kerschhaggl}, {Kh{\'e}lifi}, {Komin}, {Konopelko}, {Kosack}, {Lamanna},
  {Latham}, {Le Gallou}, {Lemi{\`e}re}, {Lemoine-Goumard}, {Lohse}, {Martin},
  {Martineau-Huynh}, {Marcowith}, {Masterson}, {Maurin}, {McComb}, {Moulin},
  {de Naurois}, {Nedbal}, {Nolan}, {Noutsos}, {Olive}, {Orford}, {Osborne},
  {Panter}, {Pelletier}, {Pita}, {P{\"u}hlhofer}, {Punch}, {Ranchon},
  {Raubenheimer}, {Raue}, {Rayner}, {Reimer}, {Reimer}, {Ripken}, {Rob},
  {Rolland}, {Rosier-Lees}, {Rowell}, {Sahakian}, {Santangelo}, {Saug{\'e}},
  {Schlenker}, {Schlickeiser}, {Schr{\"o}der}, {Schwanke}, {Schwarzburg},
  {Schwemmer}, {Shalchi}, {Sol}, {Spangler}, {Spanier}, {Steenkamp},
  {Stegmann}, {Superina}, {Tam}, {Tavernet}, {Terrier}, {Tluczykont}, {van
  Eldik}, {Vasileiadis}, {Venter}, {Vialle}, {Vincent}, {V{\"o}lk}, {Wagner},
  \& {Ward}}]{aha07_RXJ1713}
{Aharonian}, F., {Akhperjanian}, A.~G., {Bazer-Bachi}, A.~R., {et~al.}
  2007{\natexlab{a}}, \aap, 464, 235

\bibitem[{{Aharonian} {et~al.}(2007{\natexlab{b}}){Aharonian}, {Akhperjanian},
  {Bazer-Bachi}, {Beilicke}, {Benbow}, {Berge}, {Bernl{\"o}hr}, {Boisson},
  {Bolz}, {Borrel}, {Braun}, {Brown}, {B{\"u}hler}, {B{\"u}sching}, {Carrigan},
  {Chadwick}, {Chounet}, {Coignet}, {Cornils}, {Costamante}, {Degrange},
  {Dickinson}, {Djannati-Ata{\"\i}}, {Drury}, {Dubus}, {Egberts},
  {Emmanoulopoulos}, {Espigat}, {Feinstein}, {Ferrero}, {Fiasson}, {Filipovic},
  {Fontaine}, {Fukui}, {Funk}, {Funk}, {F{\"u}{\ss}ling}, {Gallant}, {Giebels},
  {Glicenstein}, {Goret}, {Hadjichristidis}, {Hauser}, {Hauser}, {Heinzelmann},
  {Henri}, {Hermann}, {Hinton}, {Hiraga}, {Hoffmann}, {Hofmann}, {Holleran},
  {Hoppe}, {Horns}, {Ishisaki}, {Jacholkowska}, {de Jager}, {Kendziorra},
  {Kerschhaggl}, {Kh{\'e}lifi}, {Komin}, {Konopelko}, {Kosack}, {Lamanna},
  {Latham}, {Le Gallou}, {Lemi{\`e}re}, {Lemoine-Goumard}, {Lohse}, {Martin},
  {Martineau-Huynh}, {Marcowith}, {Masterson}, {Maurin}, {McComb}, {Moulin},
  {Moriguchi}, {de Naurois}, {Nedbal}, {Nolan}, {Noutsos}, {Orford}, {Osborne},
  {Ouchrif}, {Panter}, {Pelletier}, {Pita}, {P{\"u}hlhofer}, {Punch},
  {Ranchon}, {Raubenheimer}, {Raue}, {Rayner}, {Reimer}, {Ripken}, {Rob},
  {Rolland}, {Rosier-Lees}, {Rowell}, {Sahakian}, {Santangelo}, {Saug{\'e}},
  {Schlenker}, {Schlickeiser}, {Schr{\"o}der}, {Schwanke}, {Schwarzburg},
  {Schwemmer}, {Shalchi}, {Sol}, {Spangler}, {Spanier}, {Steenkamp},
  {Stegmann}, {Superina}, {Tam}, {Tavernet}, {Terrier}, {Tluczykont}, {van
  Eldik}, {Vasileiadis}, {Venter}, {Vialle}, {Vincent}, {V{\"o}lk}, {Wagner},
  \& {Ward}}]{aha07_RXJ0852}
{Aharonian}, F., {Akhperjanian}, A.~G., {Bazer-Bachi}, A.~R., {et~al.}
  2007{\natexlab{b}}, \apj, 661, 236

\bibitem[{{Amato} \& {Blasi}(2006)}]{amb06}
{Amato}, E. \& {Blasi}, P. 2006, \mnras, 371, 1251

\bibitem[{{Badenes} {et~al.}(2003){Badenes}, {Bravo}, {Borkowski}, \&
  {Dom{\'{\i}}nguez}}]{bab03}
{Badenes}, C., {Bravo}, E., {Borkowski}, K.~J., \& {Dom{\'{\i}}nguez}, I. 2003,
  \apj, 593, 358

\bibitem[{{Badenes} {et~al.}(2007){Badenes}, {Hughes}, {Bravo}, \&
  {Langer}}]{bah07}
{Badenes}, C., {Hughes}, J.~P., {Bravo}, E., \& {Langer}, N. 2007, \apj, 662,
  472

\bibitem[{{Ballet}(2006)}]{ba06}
{Ballet}, J. 2006, Advances in Space Research, 37, 1902

\bibitem[{{Bamba} {et~al.}(2003){Bamba}, {Yamazaki}, {Ueno}, \&
  {Koyama}}]{bay03}
{Bamba}, A., {Yamazaki}, R., {Ueno}, M., \& {Koyama}, K. 2003, \apj, 589, 827

\bibitem[{{Baring}(2007)}]{ba07_NL}
{Baring}, M.~G. 2007, \apss, 307, 297

\bibitem[{{Bell}(1978)}]{be78a}
{Bell}, A.~R. 1978, \mnras, 182, 147

\bibitem[{{Berezhko}(1996)}]{be96}
{Berezhko}, E.~G. 1996, Astroparticle Physics, 5, 367

\bibitem[{{Berezhko} \& {Ellison}(1999)}]{bee99}
{Berezhko}, E.~G. \& {Ellison}, D.~C. 1999, \apj, 526, 385

\bibitem[{{Berezhko} {et~al.}(2002){Berezhko}, {Ksenofontov}, \&
  {V{\"o}lk}}]{bek02}
{Berezhko}, E.~G., {Ksenofontov}, L.~T., \& {V{\"o}lk}, H.~J. 2002, \aap, 395,
  943

\bibitem[{{Berezhko} {et~al.}(2003){Berezhko}, {Ksenofontov}, \&
  {V{\"o}lk}}]{bek03}
{Berezhko}, E.~G., {Ksenofontov}, L.~T., \& {V{\"o}lk}, H.~J. 2003, \aap, 412,
  L11

\bibitem[{{Berezhko} \& {V{\"o}lk}(2006)}]{bev06_RXJ1713}
{Berezhko}, E.~G. \& {V{\"o}lk}, H.~J. 2006, \aap, 451, 981

\bibitem[{{Berezhko} \& {V{\"o}lk}(2007)}]{bev07}
{Berezhko}, E.~G. \& {V{\"o}lk}, H.~J. 2007, \apjl, 661, L175

\bibitem[{{Blandford} \& {Ostriker}(1978)}]{blo78}
{Blandford}, R.~D. \& {Ostriker}, J.~P. 1978, \apjl, 221, L29

\bibitem[{{Blasi}(2002)}]{bl02}
{Blasi}, P. 2002, Astroparticle Physics, 16, 429

\bibitem[{{Blasi} {et~al.}(2005){Blasi}, {Gabici}, \& {Vannoni}}]{blg05}
{Blasi}, P., {Gabici}, S., \& {Vannoni}, G. 2005, \mnras, 361, 907

\bibitem[{{Blondin} \& {Ellison}(2001)}]{ble01}
{Blondin}, J.~M. \& {Ellison}, D.~C. 2001, \apj, 560, 244

\bibitem[{{Bykov} \& {Toptygin}(2005)}]{byt05}
{Bykov}, A.~M. \& {Toptygin}, I.~N. 2005, Astronomy Letters, 31, 748

\bibitem[{{Cassam-Chena{\"\i}} {et~al.}(2005){Cassam-Chena{\"\i}},
  {Decourchelle}, {Ballet}, \& {Ellison}}]{cad05}
{Cassam-Chena{\"\i}}, G., {Decourchelle}, A., {Ballet}, J., \& {Ellison}, D.~C.
  2005, A\&A, 443, 955

\bibitem[{Cassam-Chena{\"{i}} {et~al.}(2004)Cassam-Chena{\"{i}}, Decourchelle,
  Ballet, Sauvageot, Dubner, \& Giacani}]{cad04b}
Cassam-Chena{\"{i}}, G., Decourchelle, A., Ballet, J., {et~al.} 2004, A\&A,
  427, 199

\bibitem[{{Cassam-Chena{\"\i}} {et~al.}(2007){Cassam-Chena{\"\i}}, {Hughes},
  {Ballet}, \& {Decourchelle}}]{cah07}
{Cassam-Chena{\"\i}}, G., {Hughes}, J.~P., {Ballet}, J., \& {Decourchelle}, A.
  2007, ApJ, 665, 315

\bibitem[{{Chevalier} {et~al.}(1992){Chevalier}, {Blondin}, \&
  {Emmering}}]{chb92}
{Chevalier}, R.~A., {Blondin}, J.~M., \& {Emmering}, R.~T. 1992, \apj, 392, 118

\bibitem[{{Decourchelle}(2005)}]{de05}
{Decourchelle}, A. 2005, in X-Ray and Radio Connections (eds. L.O. Sjouwerman
  and K.K Dyer) Published electronically by NRAO,
  http://www.aoc.nrao.edu/events/xraydio Held 3-6 February 2004 in Santa Fe,
  New Mexico, USA, (E4.02) 10 pages

\bibitem[{Decourchelle {et~al.}(2000)Decourchelle, Ellison, \& Ballet}]{dee00}
Decourchelle, A., Ellison, D.~C., \& Ballet, J. 2000, ApJ, 543, L57

\bibitem[{{Dubner} {et~al.}(2002){Dubner}, {Giacani}, {Goss}, {Green}, \&
  {Nyman}}]{dug02}
{Dubner}, G.~M., {Giacani}, E.~B., {Goss}, W.~M., {Green}, A.~J., \& {Nyman},
  L.-{\AA}. 2002, \aap, 387, 1047

\bibitem[{{Dwarkadas}(2000)}]{dw00}
{Dwarkadas}, V.~V. 2000, \apj, 541, 418

\bibitem[{{Dwarkadas} \& {Chevalier}(1998)}]{dwc98}
{Dwarkadas}, V.~V. \& {Chevalier}, R.~A. 1998, \apj, 497, 807

\bibitem[{{Ellison} {et~al.}(1995){Ellison}, {Baring}, \& {Jones}}]{elb95}
{Ellison}, D.~C., {Baring}, M.~G., \& {Jones}, F.~C. 1995, \apj, 453, 873

\bibitem[{{Ellison} {et~al.}(2000){Ellison}, {Berezhko}, \& {Baring}}]{elb00}
{Ellison}, D.~C., {Berezhko}, E.~G., \& {Baring}, M.~G. 2000, \apj, 540, 292

\bibitem[{{Ellison} \& {Cassam-Chena{\"\i}}(2005)}]{elc05}
{Ellison}, D.~C. \& {Cassam-Chena{\"\i}}, G. 2005, ApJ, 632, 920

\bibitem[{Ellison {et~al.}(2004)Ellison, Decourchelle, \& Ballet}]{eld04}
Ellison, D.~C., Decourchelle, A., \& Ballet, J. 2004, A\&A, 413, 189

\bibitem[{{Ferri{\`e}re}(2001)}]{fe01_ISM}
{Ferri{\`e}re}, K.~M. 2001, Reviews of Modern Physics, 73, 1031

\bibitem[{{Ghavamian} {et~al.}(2002){Ghavamian}, {Winkler}, {Raymond}, \&
  {Long}}]{ghw02_SN1006}
{Ghavamian}, P., {Winkler}, P.~F., {Raymond}, J.~C., \& {Long}, K.~S. 2002,
  \apj, 572, 888

\bibitem[{{Hamilton} {et~al.}(2007){Hamilton}, {Fesen}, \& {Blair}}]{haf07}
{Hamilton}, A.~J.~S., {Fesen}, R.~A., \& {Blair}, W.~P. 2007, \mnras, 381, 771

\bibitem[{{Heng} \& {McCray}(2007)}]{hem07}
{Heng}, K. \& {McCray}, R. 2007, \apj, 654, 923

\bibitem[{{Hoppe} {et~al.}(2007){Hoppe}, {Lemoine-Goumard}, \& {for the
  H.~E.~S.~S.~Collaboration}}]{hol07_RCW86}
{Hoppe}, S., {Lemoine-Goumard}, M., \& {for the H.~E.~S.~S.~Collaboration}.
  2007, ArXiv e-prints, 709

\bibitem[{{Jones} \& {Ellison}(1991)}]{joe91}
{Jones}, F.~C. \& {Ellison}, D.~C. 1991, Space Science Reviews, 58, 259

\bibitem[{{Kane} {et~al.}(2000){Kane}, {Arnett}, {Remington}, {Glendinning},
  {Baz{\'a}n}, {M{\"u}ller}, {Fryxell}, \& {Teyssier}}]{kaa00}
{Kane}, J., {Arnett}, D., {Remington}, B.~A., {et~al.} 2000, \apj, 528, 989

\bibitem[{{Katz} \& {Waxman}(2008)}]{kaw08}
{Katz}, B. \& {Waxman}, E. 2008, Journal of Cosmology and Astro-Particle
  Physics, 1, 18

\bibitem[{{Korreck} {et~al.}(2004){Korreck}, {Raymond}, {Zurbuchen}, \&
  {Ghavamian}}]{kor04}
{Korreck}, K.~E., {Raymond}, J.~C., {Zurbuchen}, T.~H., \& {Ghavamian}, P.
  2004, \apj, 615, 280

\bibitem[{{Ksenofontov} {et~al.}(2005){Ksenofontov}, {Berezhko}, \&
  {V{\"o}lk}}]{ksb05}
{Ksenofontov}, L.~T., {Berezhko}, E.~G., \& {V{\"o}lk}, H.~J. 2005, \aap, 443,
  973

\bibitem[{{Leonard} {et~al.}(2005){Leonard}, {Li}, {Filippenko}, {Foley}, \&
  {Chornock}}]{lel05}
{Leonard}, D.~C., {Li}, W., {Filippenko}, A.~V., {Foley}, R.~J., \& {Chornock},
  R. 2005, \apj, 632, 450

\bibitem[{{Long} {et~al.}(2003){Long}, {Reynolds}, {Raymond}, {Winkler},
  {Dyer}, \& {Petre}}]{lor03}
{Long}, K.~S., {Reynolds}, S.~P., {Raymond}, J.~C., {et~al.} 2003, \apj, 586,
  1162

\bibitem[{{Malkov} \& {O'C Drury}(2001)}]{mad01}
{Malkov}, M.~A. \& {O'C Drury}, L. 2001, Reports of Progress in Physics, 64,
  429

\bibitem[{{Matzner} \& {McKee}(1999)}]{mam99}
{Matzner}, C.~D. \& {McKee}, C.~F. 1999, \apj, 510, 379

\bibitem[{{Moffett} {et~al.}(2004){Moffett}, {Caldwell}, {Reynoso}, \&
  {Hughes}}]{moc04}
{Moffett}, D., {Caldwell}, C., {Reynoso}, E., \& {Hughes}, J. 2004, in IAU
  Symposium, Vol. 218, Young Neutron Stars and Their Environments, ed.
  F.~{Camilo} \& B.~M. {Gaensler}, 69--+

\bibitem[{{Moffett} {et~al.}(1993){Moffett}, {Goss}, \& {Reynolds}}]{mog93}
{Moffett}, D.~A., {Goss}, W.~M., \& {Reynolds}, S.~P. 1993, \aj, 106, 1566

\bibitem[{{Nomoto} {et~al.}(1984){Nomoto}, {Thielemann}, \& {Yokoi}}]{not84}
{Nomoto}, K., {Thielemann}, F.-K., \& {Yokoi}, K. 1984, \apj, 286, 644

\bibitem[{{O'C Drury} {et~al.}(1996){O'C Drury}, {Duffy}, \& {Kirk}}]{drd96}
{O'C Drury}, L., {Duffy}, P., \& {Kirk}, J.~G. 1996, \aap, 309, 1002

\bibitem[{{Orlando} {et~al.}(2007){Orlando}, {Bocchino}, {Reale}, {Peres}, \&
  {Petruk}}]{orb07}
{Orlando}, S., {Bocchino}, F., {Reale}, F., {Peres}, G., \& {Petruk}, O. 2007,
  \aap, 470, 927

\bibitem[{{Parizot} {et~al.}(2006){Parizot}, {Marcowith}, {Ballet}, \&
  {Gallant}}]{pam06}
{Parizot}, E., {Marcowith}, A., {Ballet}, J., \& {Gallant}, Y.~A. 2006, \aap,
  453, 387

\bibitem[{{Plaga}(2008)}]{pl08}
{Plaga}, R. 2008, New Astronomy, 13, 73

\bibitem[{{Porter} {et~al.}(2006){Porter}, {Moskalenko}, \& {Strong}}]{pom06}
{Porter}, T.~A., {Moskalenko}, I.~V., \& {Strong}, A.~W. 2006, \apjl, 648, L29

\bibitem[{{Raymond} {et~al.}(1995){Raymond}, {Blair}, \& {Long}}]{rab95}
{Raymond}, J.~C., {Blair}, W.~P., \& {Long}, K.~S. 1995, \apjl, 454, L31+

\bibitem[{{Raymond} {et~al.}(2007){Raymond}, {Korreck}, {Sedlacek}, {Blair},
  {Ghavamian}, \& {Sankrit}}]{rak07}
{Raymond}, J.~C., {Korreck}, K.~E., {Sedlacek}, Q.~C., {et~al.} 2007, \apj,
  659, 1257

\bibitem[{{Reynolds}(1998)}]{re98}
{Reynolds}, S.~P. 1998, \apj, 493, 375

\bibitem[{{Reynolds} \& {Gilmore}(1986)}]{reg86}
{Reynolds}, S.~P. \& {Gilmore}, D.~M. 1986, \aj, 92, 1138

\bibitem[{{Reynolds} \& {Gilmore}(1993)}]{reg93}
{Reynolds}, S.~P. \& {Gilmore}, D.~M. 1993, \aj, 106, 272

\bibitem[{{Reynoso} {et~al.}(2008){Reynoso}, {Castelletti}, {Moffett}, \&
  {Hughes}}]{rec08}
{Reynoso}, E.~M., {Castelletti}, G.~M., {Moffett}, D.~A., \& {Hughes}, J.~P.
  2008, RMxAAC, 27

\bibitem[{{Rothenflug} {et~al.}(2004){Rothenflug}, {Ballet}, {Dubner},
  {Giacani}, {Decourchelle}, \& {Ferrando}}]{rob04}
{Rothenflug}, R., {Ballet}, J., {Dubner}, G., {et~al.} 2004, \aap, 425, 121

\bibitem[{{Sollerman} {et~al.}(2003){Sollerman}, {Ghavamian}, {Lundqvist}, \&
  {Smith}}]{sog03}
{Sollerman}, J., {Ghavamian}, P., {Lundqvist}, P., \& {Smith}, R.~C. 2003,
  \aap, 407, 249

\bibitem[{{Uchiyama} {et~al.}(2007){Uchiyama}, {Aharonian}, {Tanaka},
  {Takahashi}, \& {Maeda}}]{uca07}
{Uchiyama}, Y., {Aharonian}, F.~A., {Tanaka}, T., {Takahashi}, T., \& {Maeda},
  Y. 2007, \nat, 449, 576

\bibitem[{{Vink} {et~al.}(2003){Vink}, {Laming}, {Gu}, {Rasmussen}, \&
  {Kaastra}}]{vil03b_SN1006}
{Vink}, J., {Laming}, J.~M., {Gu}, M.~F., {Rasmussen}, A., \& {Kaastra}, J.~S.
  2003, \apjl, 587, L31

\bibitem[{{V{\"o}lk} {et~al.}(2003){V{\"o}lk}, {Berezhko}, \&
  {Ksenofontov}}]{vob03}
{V{\"o}lk}, H.~J., {Berezhko}, E.~G., \& {Ksenofontov}, L.~T. 2003, \aap, 409,
  563

\bibitem[{{Wang} \& {Chevalier}(2001)}]{wac01}
{Wang}, C.-Y. \& {Chevalier}, R.~A. 2001, \apj, 549, 1119

\bibitem[{{Warren} {et~al.}(2005){Warren}, {Hughes}, {Badenes}, {Ghavamian},
  {McKee}, {Moffett}, {Plucinsky}, {Rakowski}, {Reynoso}, \& {Slane}}]{wah05}
{Warren}, J.~S., {Hughes}, J.~P., {Badenes}, C., {et~al.} 2005, ApJ, 634, 376

\bibitem[{{Winkler} {et~al.}(2003){Winkler}, {Gupta}, \& {Long}}]{wig03}
{Winkler}, P.~F., {Gupta}, G., \& {Long}, K.~S. 2003, \apj, 585, 324

\bibitem[{{Winkler} \& {Long}(1997)}]{wil97b_SN1006}
{Winkler}, P.~F. \& {Long}, K.~S. 1997, \apj, 491, 829

\bibitem[{{Winkler} {et~al.}(2005){Winkler}, {Long}, {Hamilton}, \&
  {Fesen}}]{wil05}
{Winkler}, P.~F., {Long}, K.~S., {Hamilton}, A.~J.~S., \& {Fesen}, R.~A. 2005,
  \apj, 624, 189

\bibitem[{{Yamaguchi} {et~al.}(2007){Yamaguchi}, {Koyama}, {Katsuda},
  {Nakajima}, {Hughes}, {Bamba}, {Hiraga}, {Mori}, {Ozaki}, \& {Tsuru}}]{yak07}
{Yamaguchi}, H., {Koyama}, K., {Katsuda}, S., {et~al.} 2007, ArXiv e-prints,
  706

\end{thebibliography}





\clearpage

\begin{figure*}[t] 
\centering
\includegraphics[width=14cm]{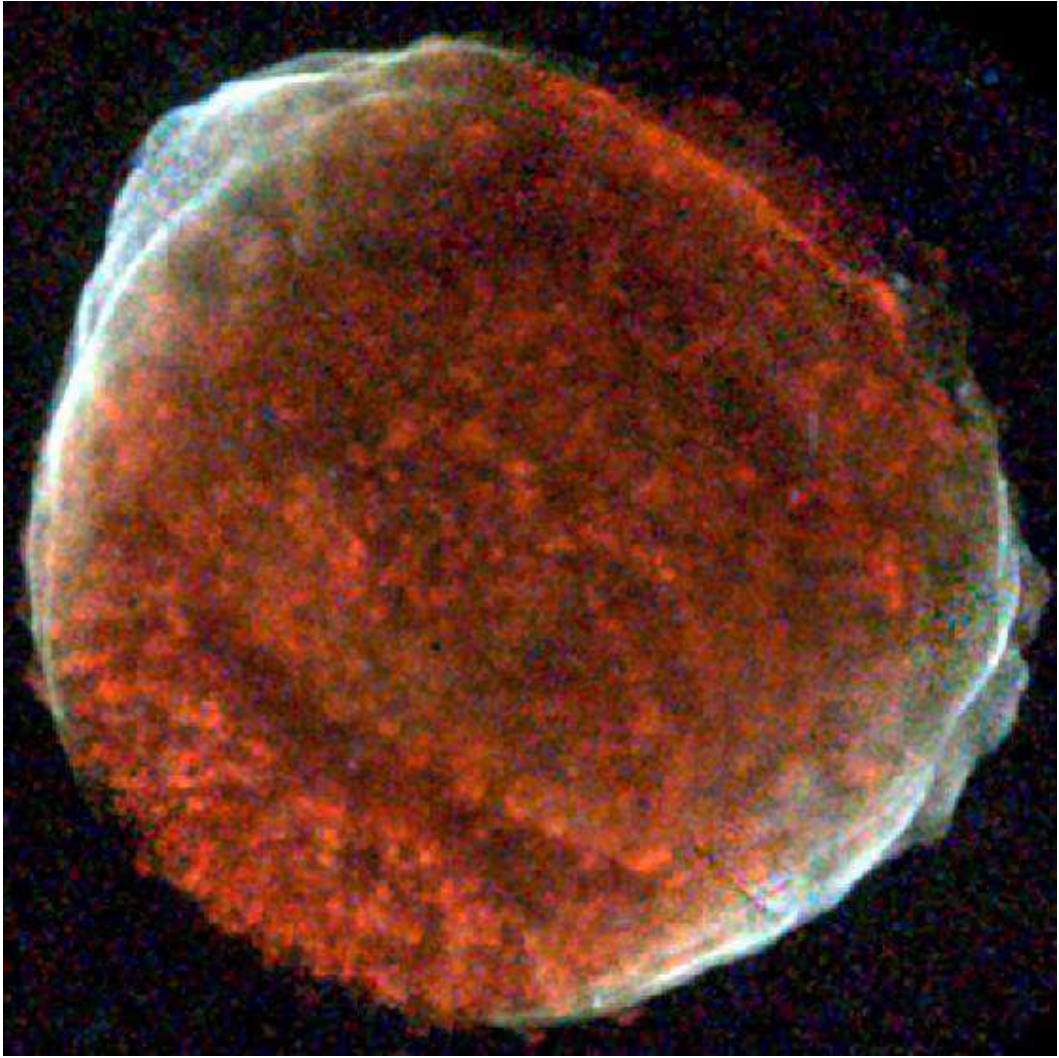}
\caption{Three-color \textit{Chandra} image of SN 1006: 0.5-0.8 keV
(\textit{red}), 0.8-2 keV (\textit{green}) and 2-4.5 keV
(\textit{blue}). The regions dominated by synchrotron emission
from high-energy relativistic electrons appears in white and are
naturally associated with the BW. The regions dominated by the
thermal emission from the shocked gas (mostly oxygen) appears in red
and are most likely associated with the ejecta. Note that the oxygen
is found slightly ahead of the BW in some places (e.g., east
and south).  Point sources have been removed here so that only the
diffuse emission from the remnant is visible. Images were background
subtracted, corrected for vignetting and slightly smoothed.  The
intensity scaling is square-root with a maximum fixed at $5.0 \times
10^{-4}$, $5.0 \times 10^{-4}$ and $2.0 \times 10^{-4}$
ph/cm$^2$/s/arcmin$^2$ for the red, green and blue bands,
respectively.} 
\label{fig_SN1006_3color_xray} 
\end{figure*}

\begin{figure*}[t] 
\centering
\includegraphics[width=14cm]{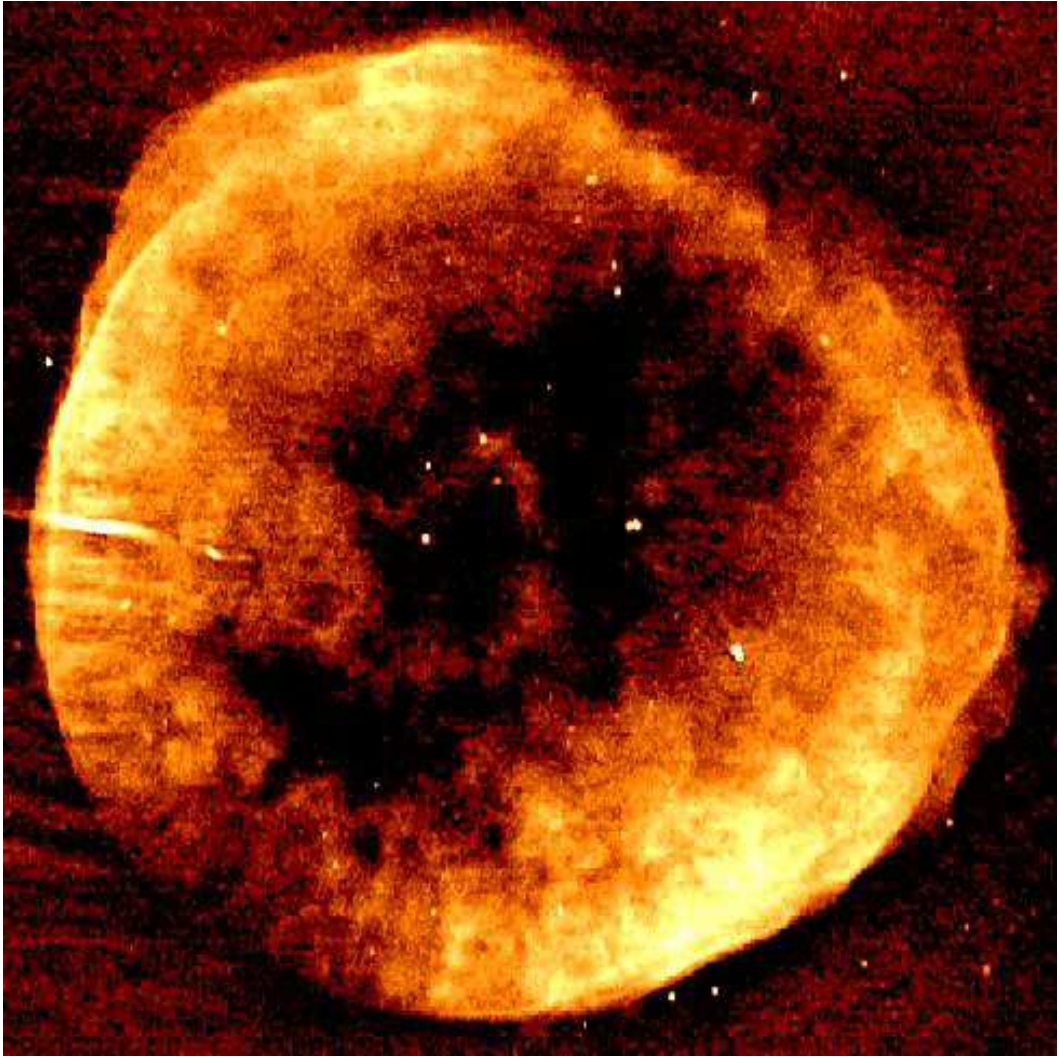}
\caption{Radio image of SN 1006 at 1.5 GHz. The
intensity scaling is square-root with a maximum fixed at $1.0$ mJy$/$beam.
The bright elongated source in the western rim is a background radio galaxy.} 
\label{fig_SN1006_radio} 
\end{figure*}

\begin{figure*}[t] 
\centering
\includegraphics[width=14cm]{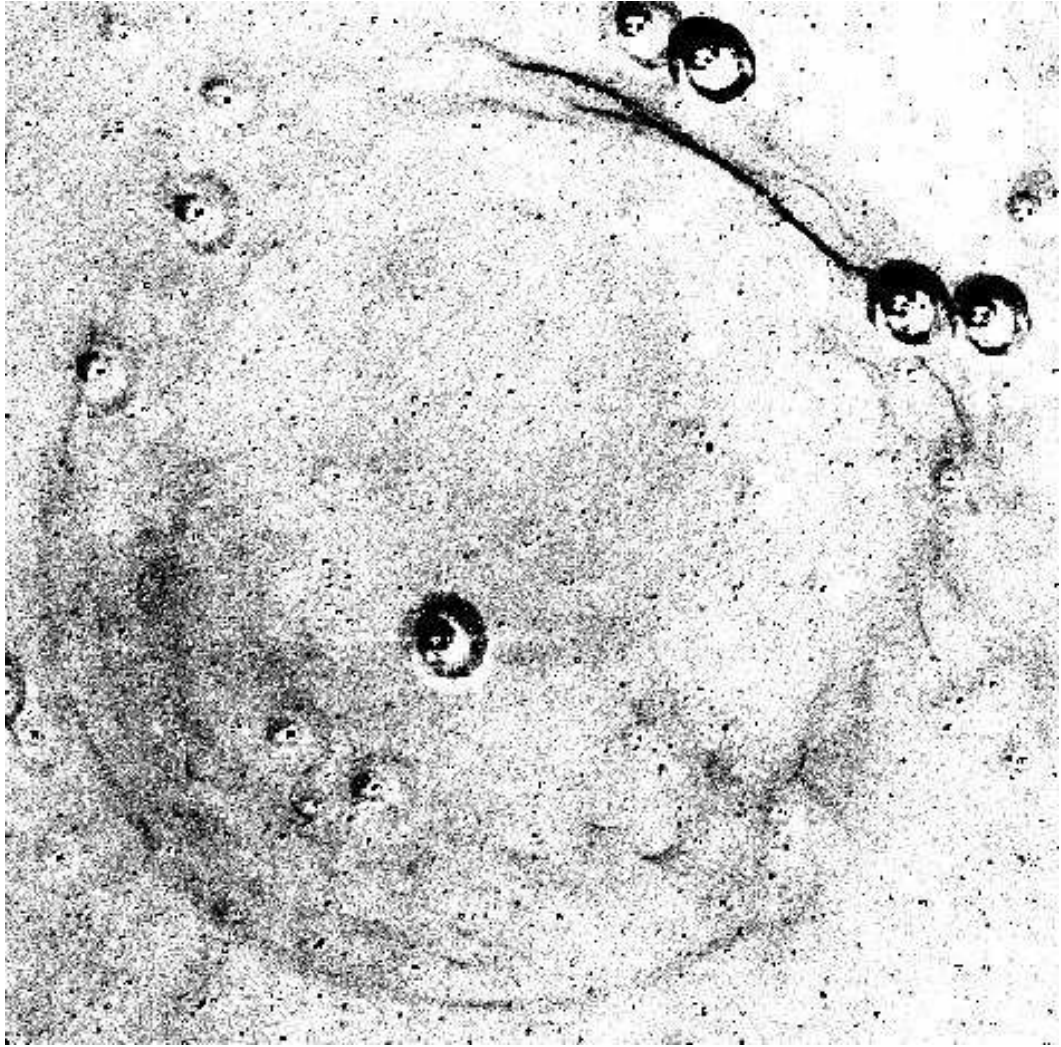}
\caption{H$\alpha$ image of SN 1006.} 
\label{fig_SN1006_Halpha} 
\end{figure*}

\begin{figure*}[t] 
\centering
\includegraphics[width=8cm]{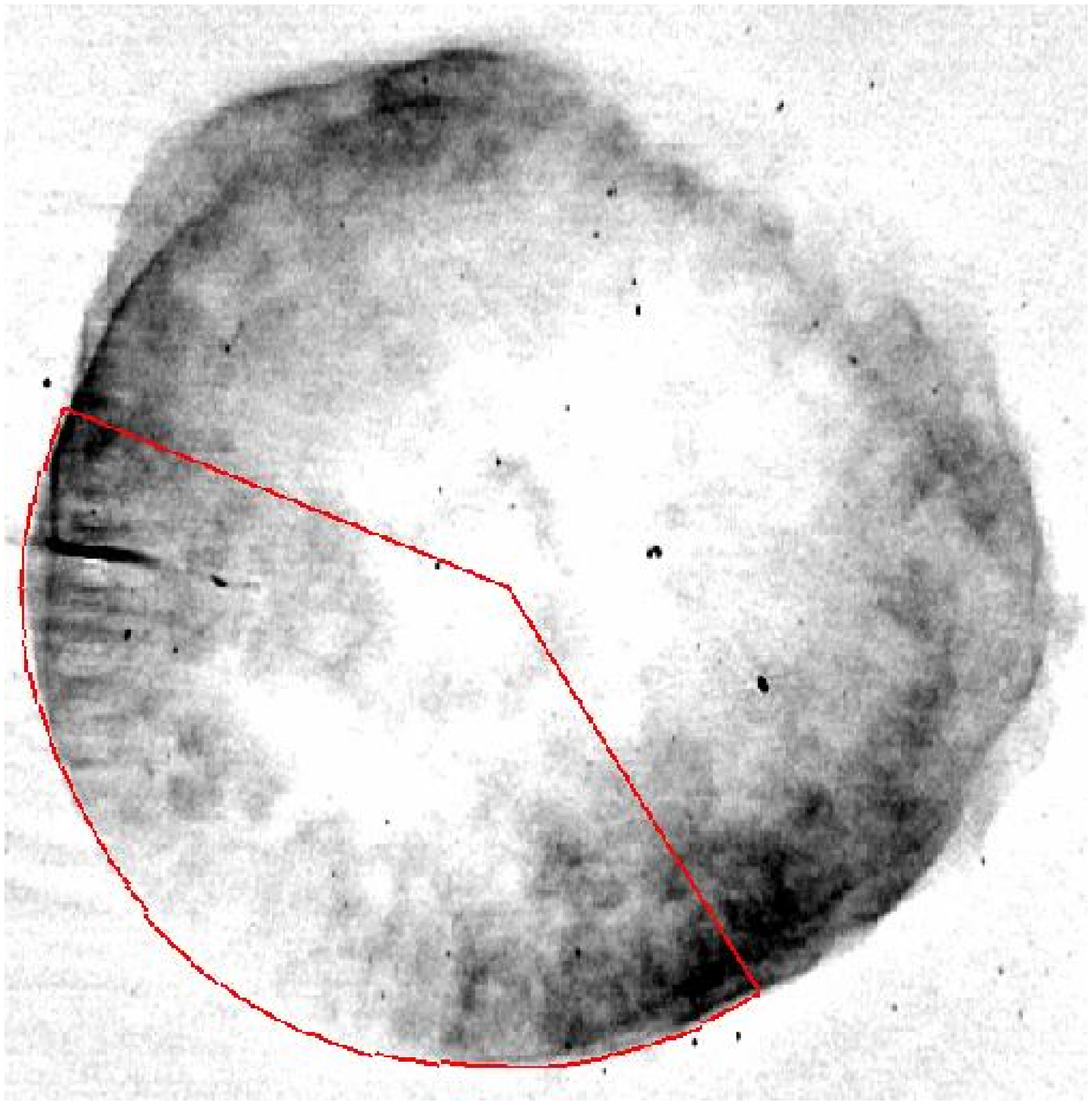}
\includegraphics[width=8cm]{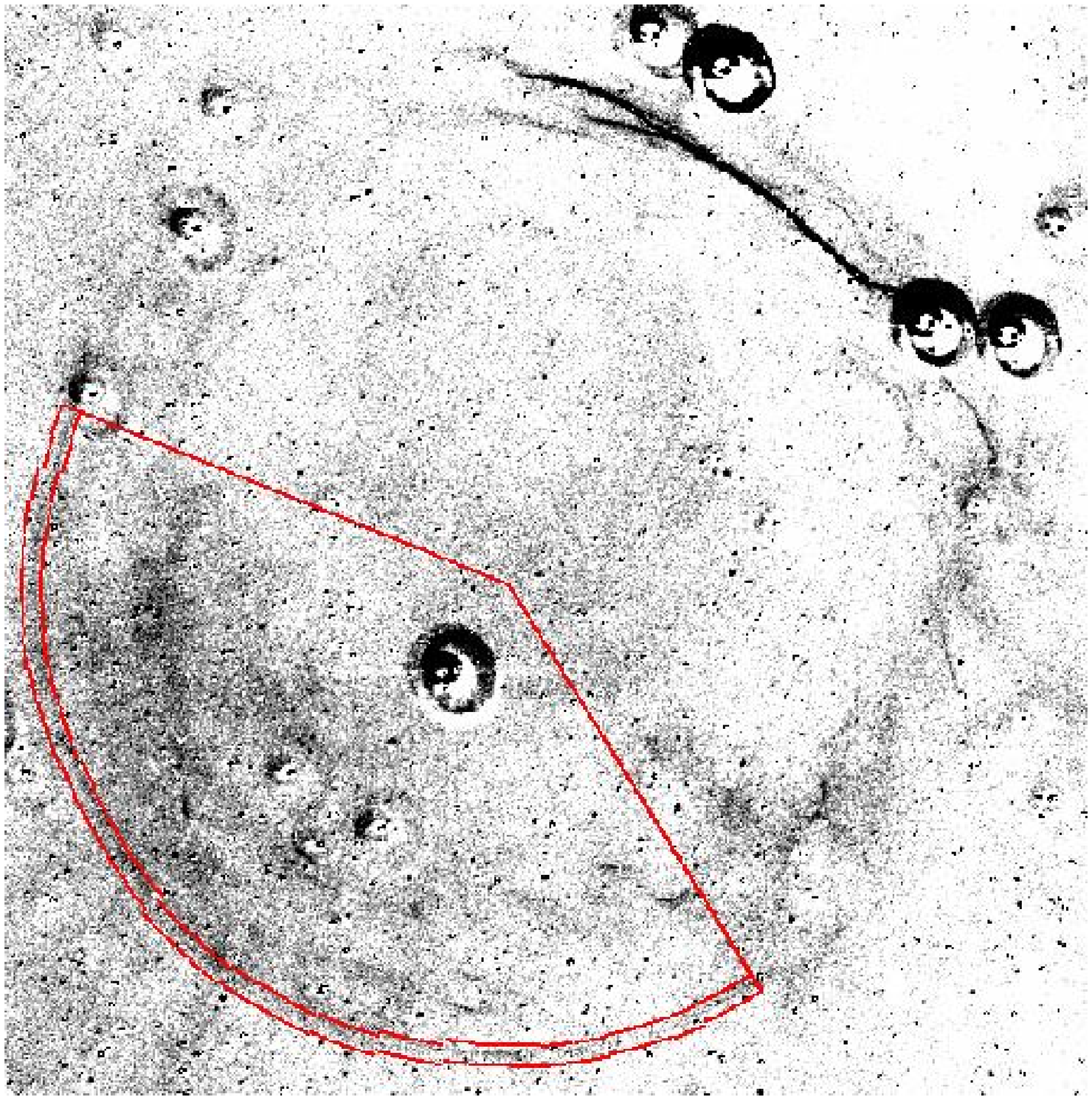}
\includegraphics[width=8cm]{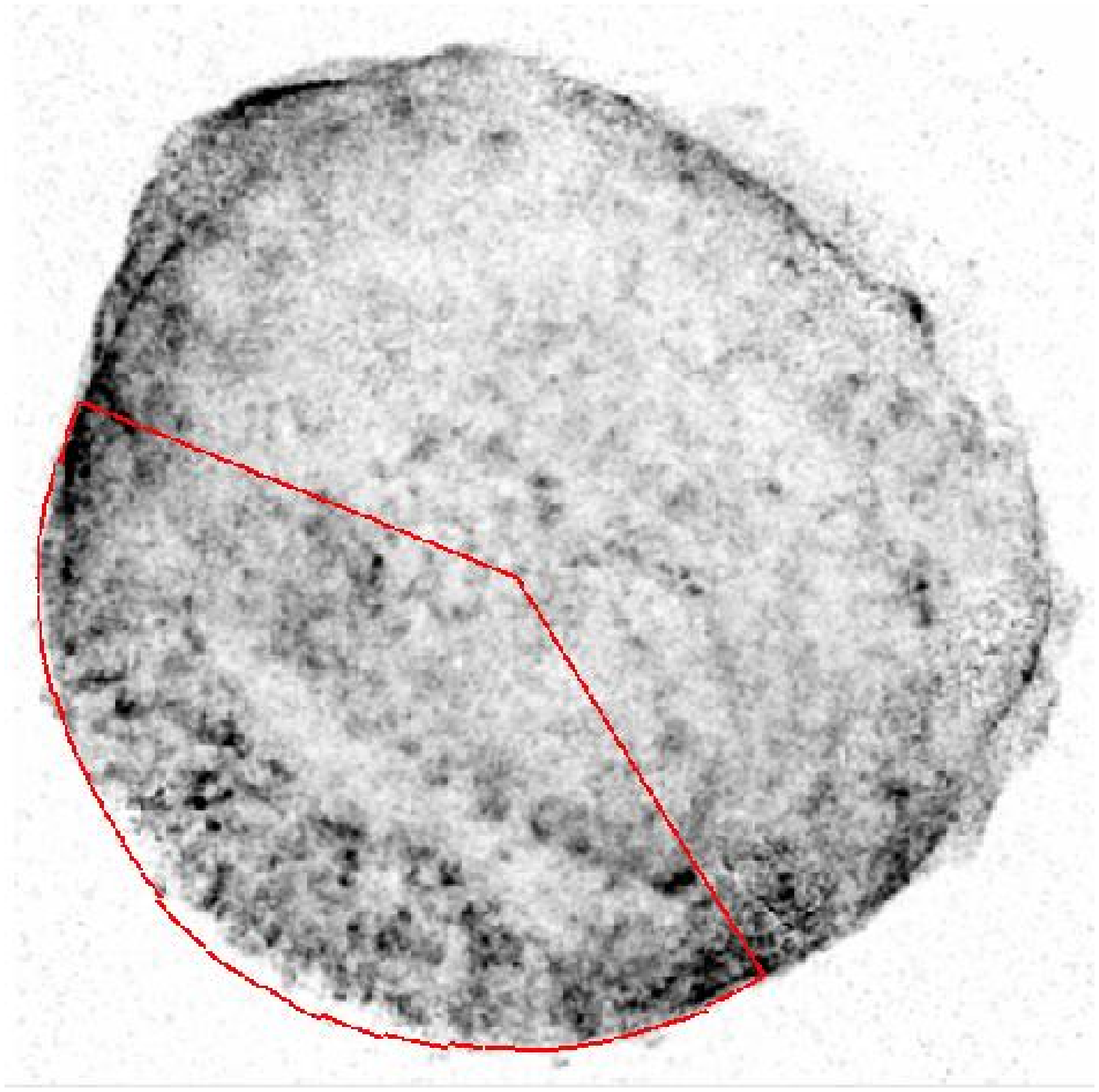}
\includegraphics[width=8cm]{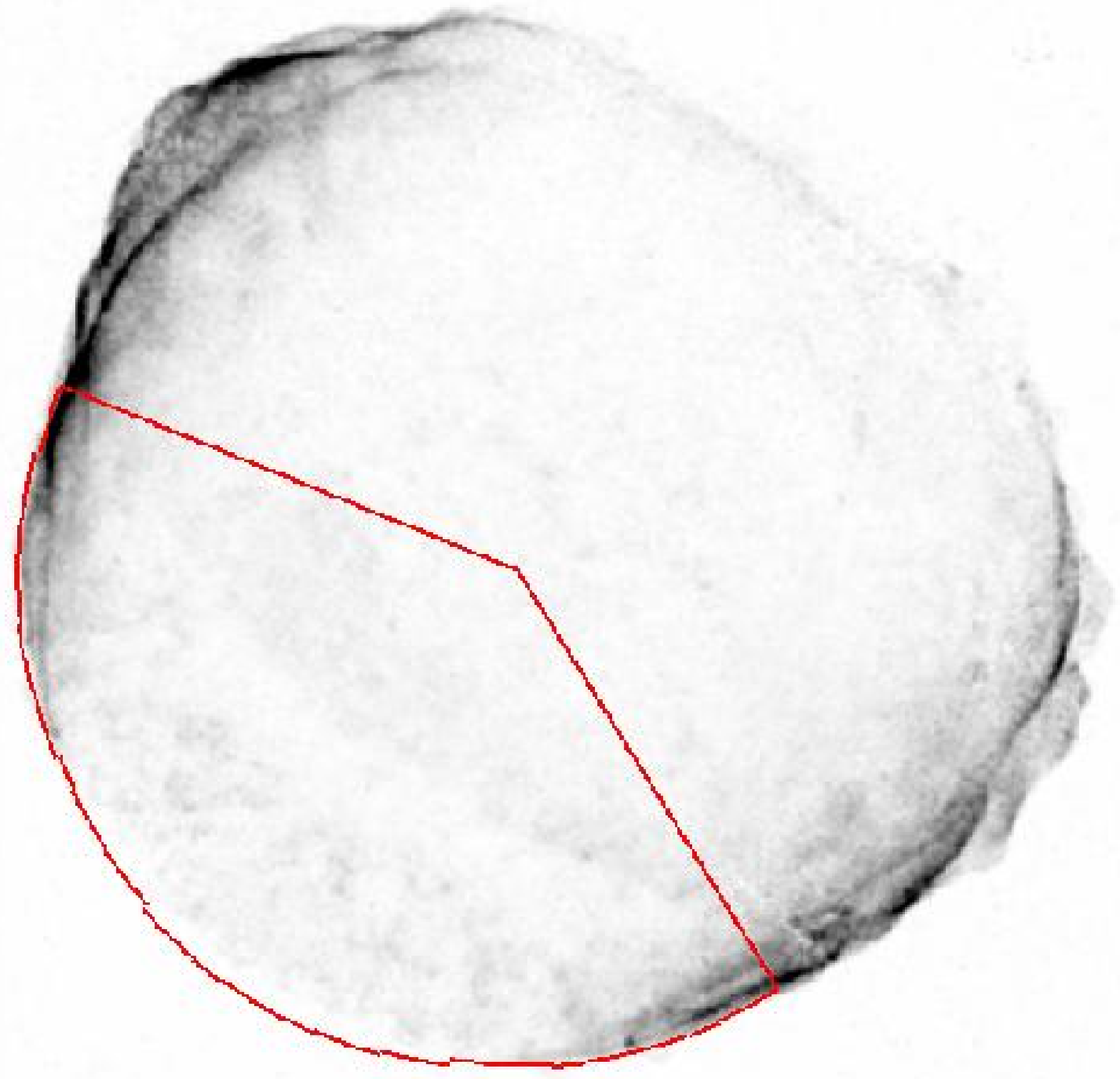}
\caption{Images of SN 1006 in several energy bands. \textit{Top-left}:
Radio image at 1.5 GHz. The maximum intensity was fixed at $0.6$
mJy$/$beam. \textit{Top-right}: H$\alpha$ image with the contour
errors associated with the shock front seen in the southeast
(\textit{red lines}). \textit{Bottom-left}: \textit{Chandra} X-ray
image in the oxygen band (0.5-0.8 keV). The maximum intensity was
fixed at $5.0 \times 10^{-4}$ ph/cm$^2$/s/arcmin$^2$.
\textit{Bottom-right}: \textit{Chandra} image in the mid-energy X-ray
band (0.8-2.0 keV). The maximum intensity was fixed at $6.0 \times
10^{-4}$ ph/cm$^2$/s/arcmin$^2$. Both X-ray images were background
subtracted, corrected for vignetting and slightly smoothed. In the
radio and X-ray images, we show the contour derived from the H$\alpha$
emission (\textit{red lines}). This contour lies within the contour
errors shown in the H$\alpha$ image. In all the images, the intensity
scaling is linear.}
\label{fig_SN1006_all} 
\end{figure*}

\begin{figure*}[t] 
\centering
\includegraphics[width=13cm]{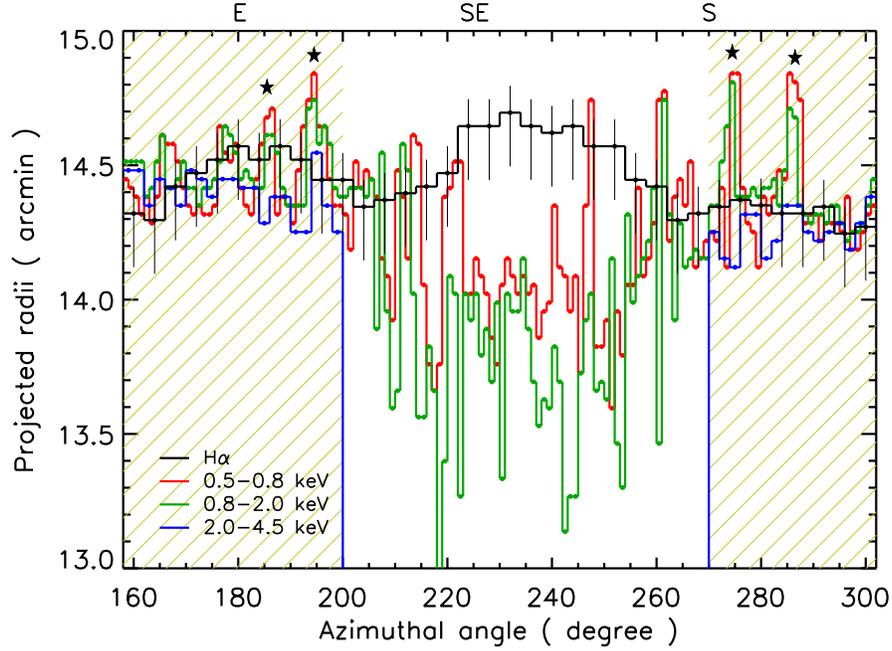} 
\caption{Azimuthal variation of the outer H$\alpha$ emission
(\textit{black lines}) and X-ray emission as obtained in several X-ray
energy bands: 0.5-0.8 keV (\textit{red lines}), 0.8-2 keV
(\textit{green lines}) and 2-4.5 keV (\textit{blue lines}). Regions
where the X-ray synchrotron emission is dominant are indicated by
cross-hatched yellow lines.  In those regions, the fingers
(\textit{stars}) indicate the presence of shocked ejecta found at or
even slightly ahead of the BW. The X-ray contours correspond
to places where the brightness becomes larger than $1.5 \times
10^{-5}$ ph/cm$^2$/s/arcmin$^2$ in the $0.5-0.8$ keV band, $0.6 \times
10^{-5}$ ph/cm$^2$/s/arcmin$^2$ in the $0.8-2$ keV band and $0.4
\times 10^{-5}$ ph/cm$^2$/s/arcmin$^2$ in the $2-4.5$ keV band. Note
that the radii measurements are quite sensitive to the contour values,
but the range of variations is contained between the red and blue
lines.}
\label{fig_radius_Ha_xray} 
\end{figure*}

\begin{figure*}[t] 
\centering
\includegraphics[width=13cm]{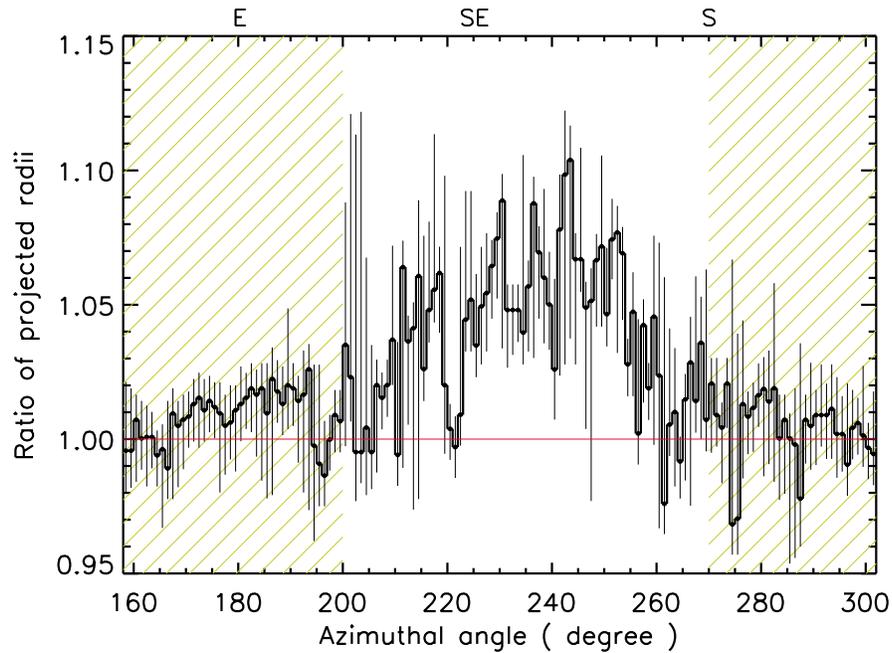} 
\caption{Ratio of the outer H$\alpha$ radius (\textit{black lines} in
Fig.  \ref{fig_radius_Ha_xray}) to the 0.5-0.8 keV X-ray radius
(\textit{red lines} in Fig. \ref{fig_radius_Ha_xray}). Regions where
the X-ray synchrotron emission is dominant are indicated by
cross-hatched yellow lines. Error bars include uncertainties on the
radius derived from the (expanded) H$\alpha$ image and that
done in the low-energy X-rays choosing different
contour values ($2.5 \pm 1.0 \times 10^{-5}$
ph/cm$^2$/s/arcmin$^2$). The large error bars between
$200^\circ-205^\circ$ are due to a low exposure there.}
\label{fig_Rs_o_Rc} 
\end{figure*}

\begin{figure*}[t] 
\centering
\includegraphics[width=13cm]{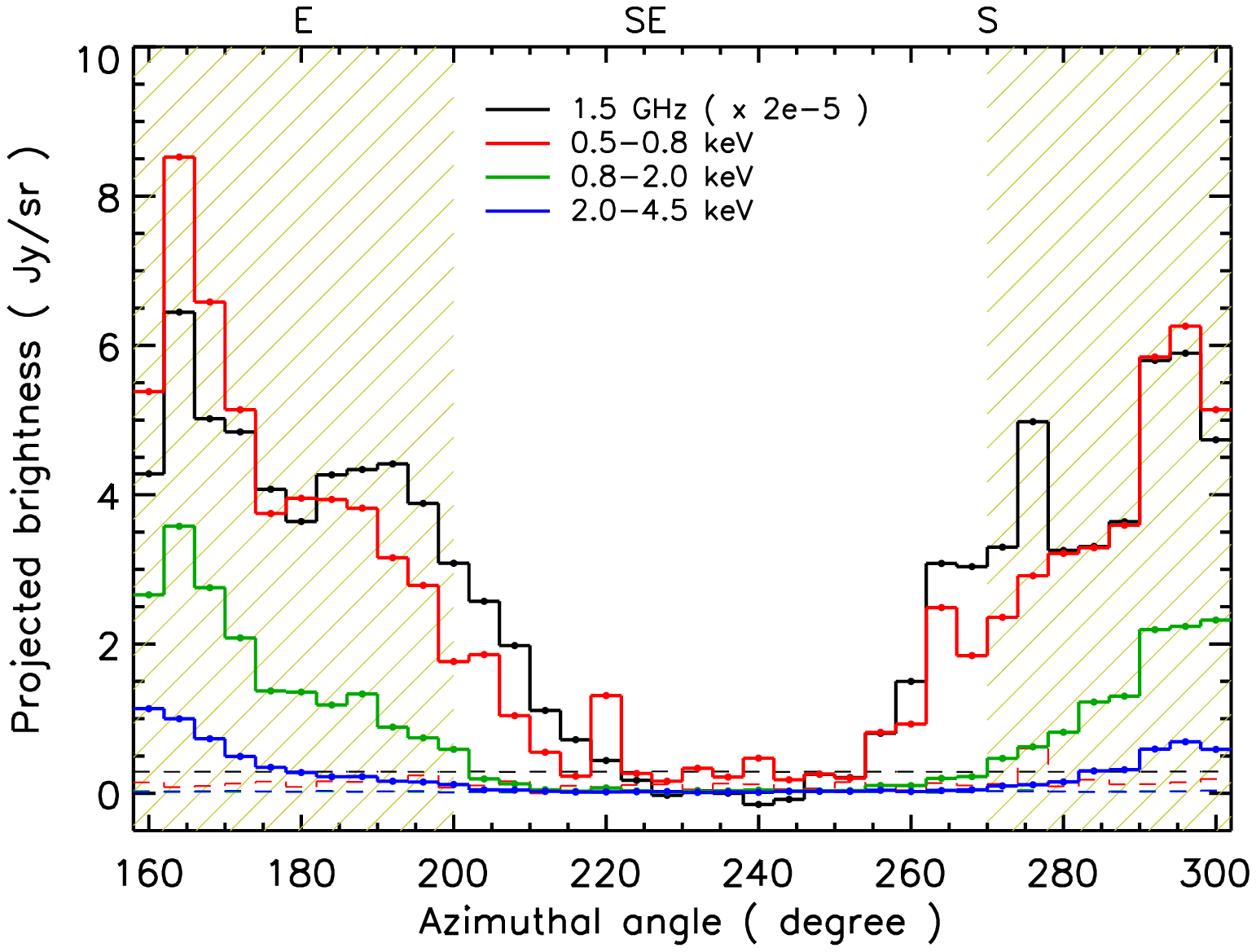}
\includegraphics[width=13cm]{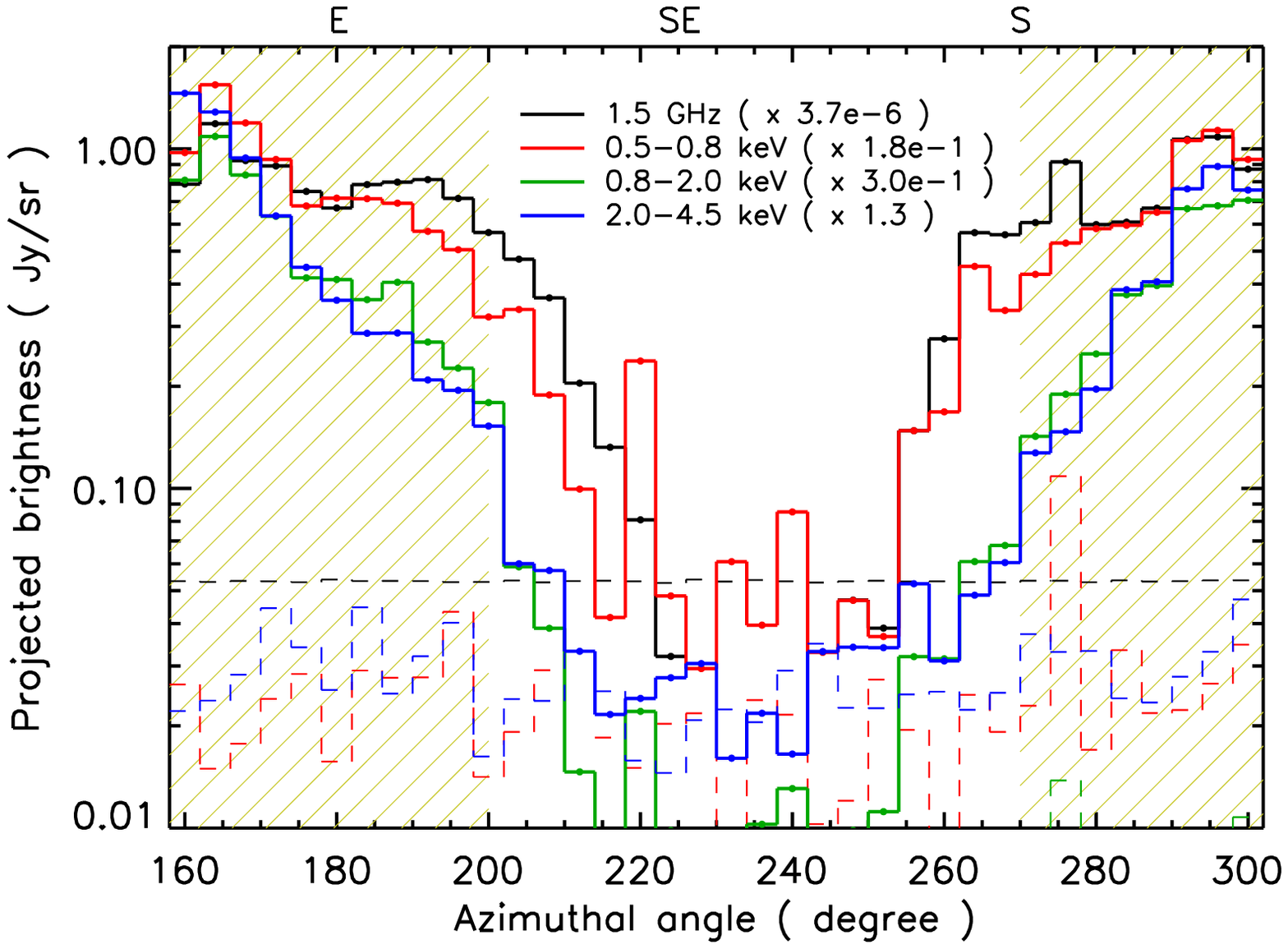}
\caption{\textit{Top}: Azimuthal variations of the brightness at the
BW of SN 1006 at several frequencies/energies: 1.5 GHz (\textit{black
lines}), 0.5-0.8 keV (\textit{red lines}), 0.8-2 keV (\textit{green
lines}) and 2-4.5 keV (\textit{blue lines}). The brightness in each
azimuthal bin was calculated between the BW radius and the radius
$30\arcsec$ behind.  For comparison, we show the level of the radio
rms noise level and X-ray background located $1\arcmin$ outside the BW
and extracted from a $30\arcsec$-wide box (\textit{dashed lines}).
\textit{Bottom}: Same but with the curves rescaled roughly to the same
level in the brightest regions and shown with a log scale.}
\label{fig_flux_vs_angle_radio_xrays} 
\end{figure*}

\begin{figure*}[t] 
\centering 
\includegraphics[width=13cm]{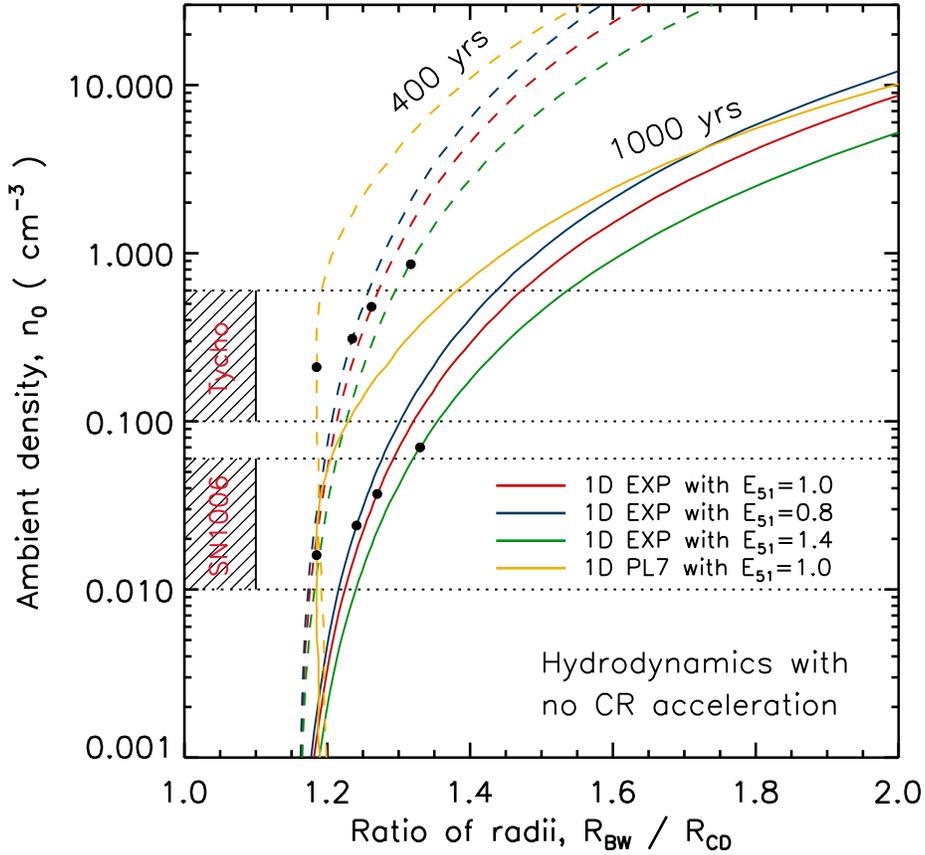} 
\caption{Ratio of radii between the BW and CD,
$R_{\mathrm{BW}}/R_{\mathrm{CD}}$, as obtained from 1-D numerical
hydrodynamical simulations (assuming no CR acceleration at the BW) and
calculated for a large range of ambient density, $n_0$, for a fixed
SNR age (\textit{dashed lines}: 400 yrs; \textit{solid lines}: 1000
yrs).  We show curves obtained with a 1-D exponential (EXP) ejecta
profile with a kinetic energy of the explosion $E_{51} \equiv E /
10^{51}$ ergs $=1.0$ (\textit{red lines}), $E_{51} = 0.8$
(\textit{blue lines}), $E_{51} = 1.4$ (\textit{green lines}) and a 1-D
powerlaw (PL7) ejecta profile ($n=7$) with $E_{51} = 1.0$
(\textit{yellow lines}). The cross-hatched domains correspond to the
intersection of the range of ratio of radii observed in SN 1006 and
Tycho ($\leq 1.10$, \textit{vertical solid lines}) and range of
ambient medium density typically encountered in their vicinity
(\textit{horizontal dotted lines}). The 1-D hydrodynamical simulations
are clearly unable to predict a ratio of radii as small as the one
observed in both SNRs. For completeness, we indicate by a
filled circle the value of the ambient density at which the simulated
SNR radius matches the one observed for each model. An angular BW
radius of $14.6\arcmin$ (resp. $256\arcsec$) and a distance to the SNR
of $2.0$ kpc ($2.8$ kpc) were assumed for SN 1006 (resp. Tycho),
resulting in a physical BW radius of $\sim 8.5$ pc (resp. $\sim 3.5$
pc). Larger values for the physical radius (due for instance to a
larger distance to the remnant) would correspond to lower ambient
densities.}
\label{fig_n0_vs_ratio} 
\end{figure*}

\begin{figure*}[t] 
\centering 
\includegraphics[width=13cm]{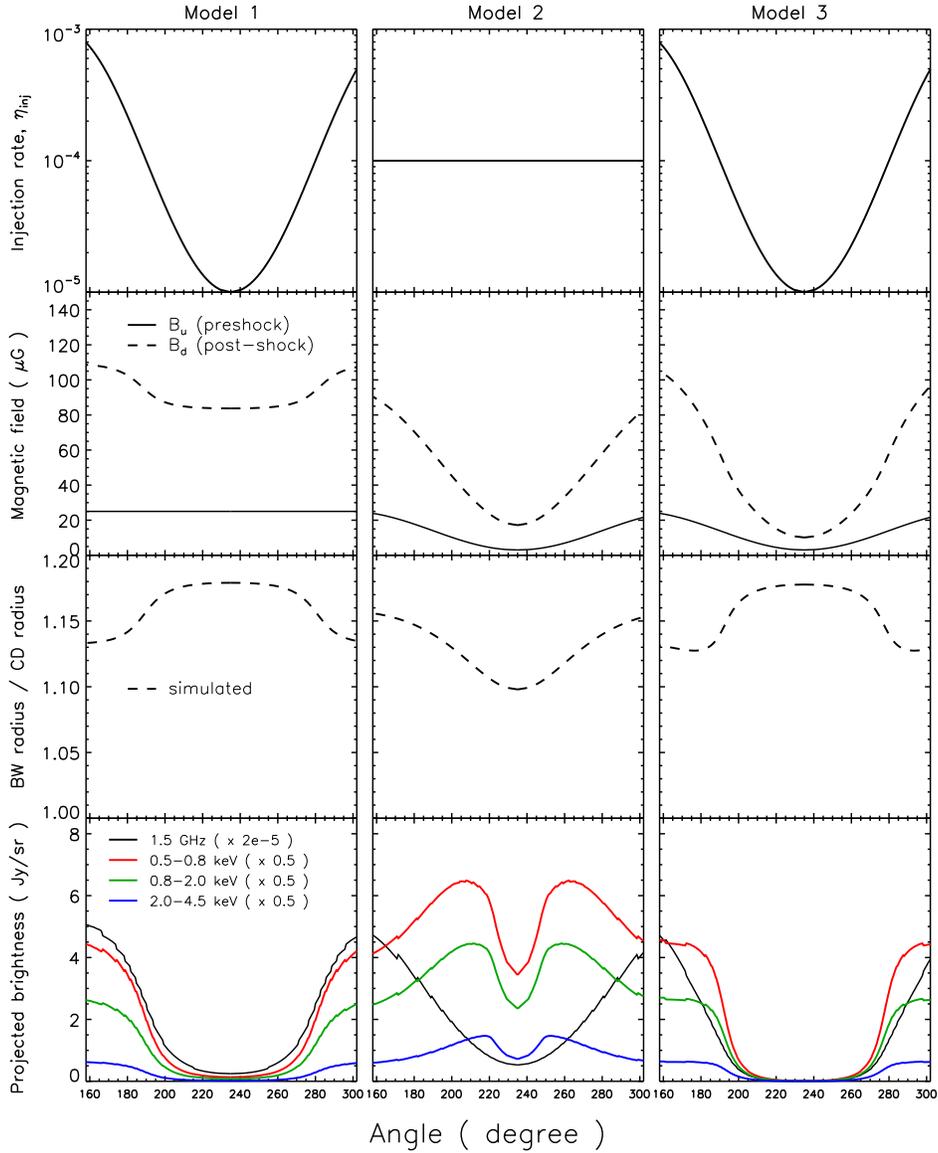} 
\caption{Various models assuming different azimuthal profiles for the
injection rate, $\eta_{\mathrm{inj}}$ (\textit{top}), and far
upstream magnetic field, $B_{\mathrm{u}}$ (\textit{second, solid lines}). Each
model predicts azimuthal variations of the ratio of radii between the
BW and CD, $R_{\mathrm{BW}}/R_{\mathrm{CD}}$ (\textit{third}), and
synchrotron brightness (projected along the line-of-sight) at different 
frequencies (\textit{bottom}).}
\label{fig_heuristic_model_profiles} 
\end{figure*}

\begin{figure*}[t] 
\centering \includegraphics[bb=0 60 720 720,clip,width=18cm]{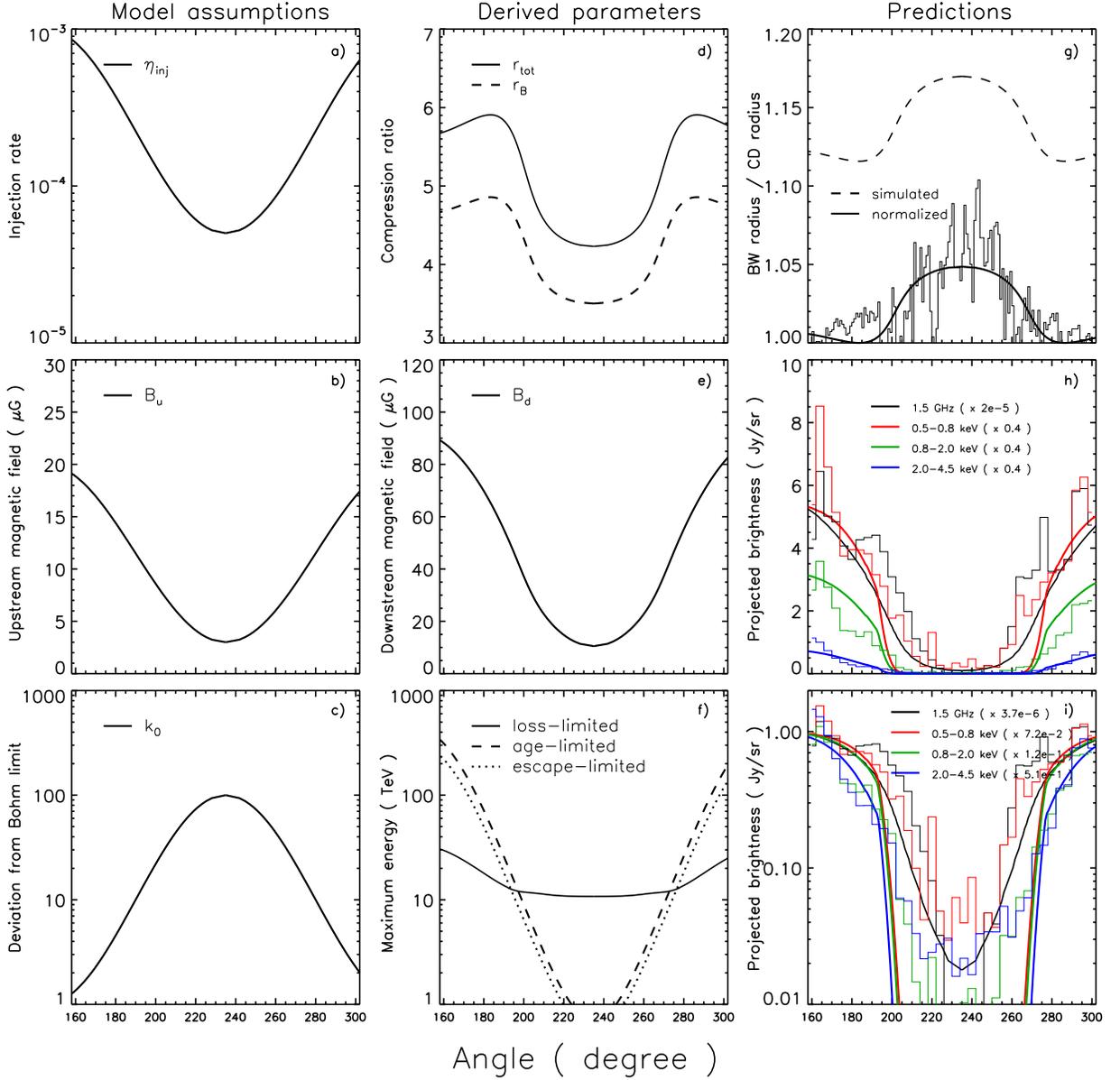}
\caption{\textit{Left column:} azimuthal profiles assumed for the
injection efficiency, $\eta_{\mathrm{inj}}$ (\textit{panel a}), far
upstream magnetic field, $B_{\mathrm{u}}$ (\textit{panel b}) and
diffusion coefficient normalized to the Bohm-limit value, $k_0$
(\textit{panel c}).  \textit{Middle column:} model profiles of the
overall density ($r_{\mathrm{tot}}$) and magnetic field ($r_{B}$)
compression ratios (\textit{panel d}), immediate post-shock value of
the magnetic field, $B_{\mathrm{d}}$ (\textit{panel e}) and maximum
energies (\textit{panel f}).  \textit{Right column:} predicted
(\textit{thick} lines) and observed (\textit{thin} histograms)
azimuthal profiles for the ratio of radii between the BW and CD,
$R_{\mathrm{BW}}/R_{\mathrm{CD}}$ (\textit{panel g, solid lines}) and
line-of-sight projected synchrotron brightness at several frequencies
(\textit{panels h and i}).}
\label{fig_model_k0_var_profiles} 
\end{figure*}

\begin{figure*}[t] 
\centering 
\includegraphics[width=6cm]{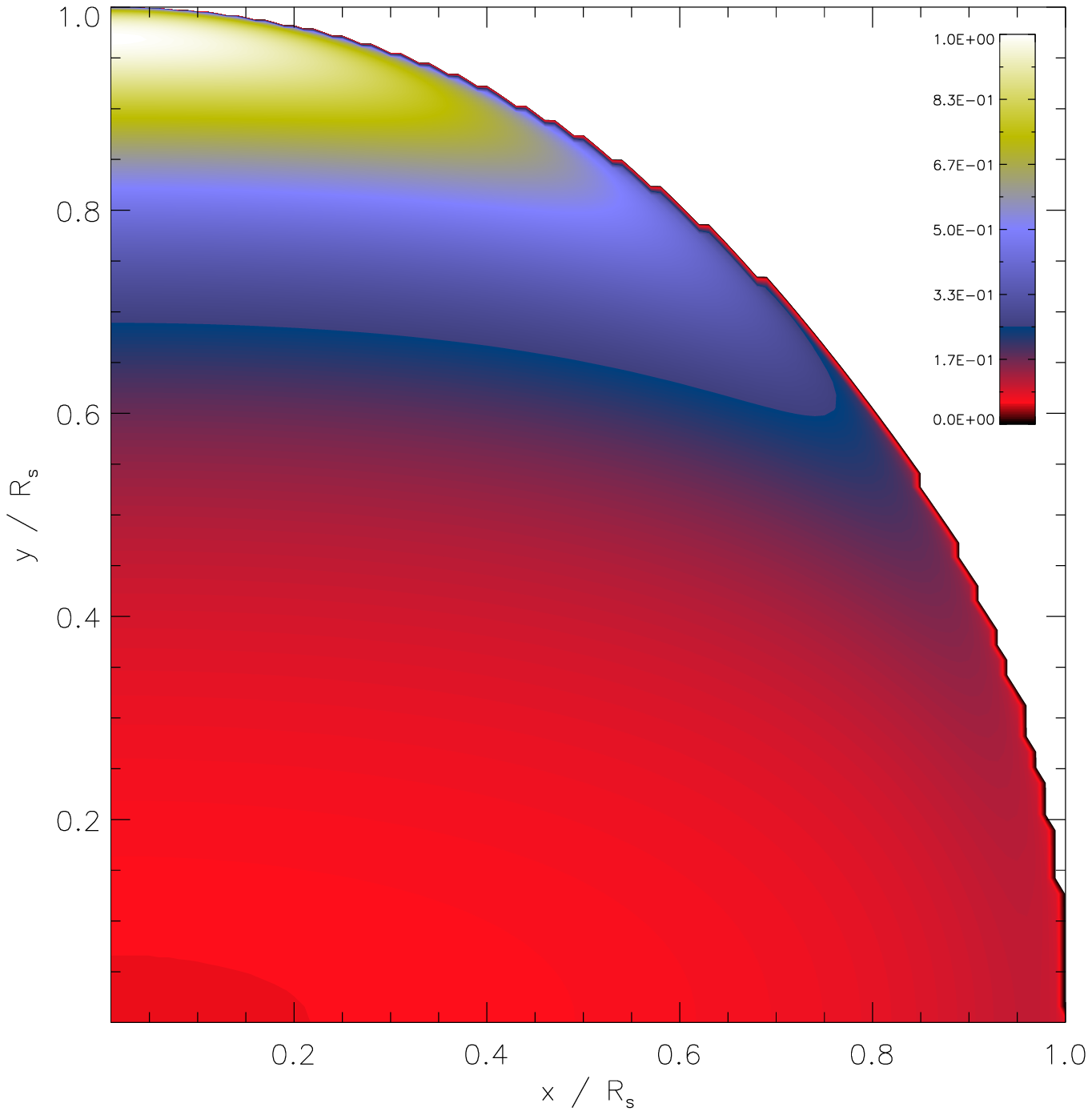} 
\includegraphics[width=6cm]{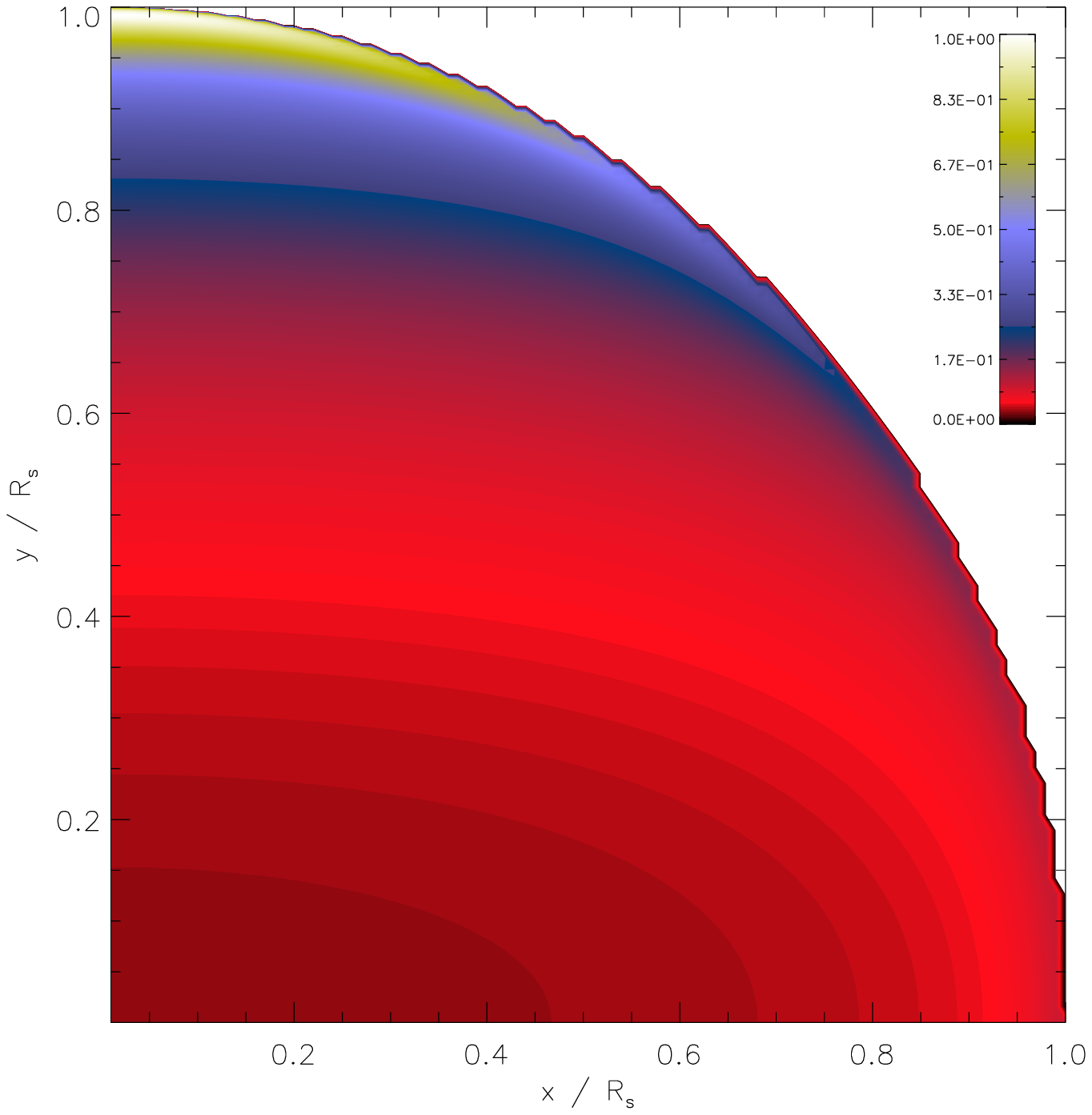} 
\includegraphics[width=6cm]{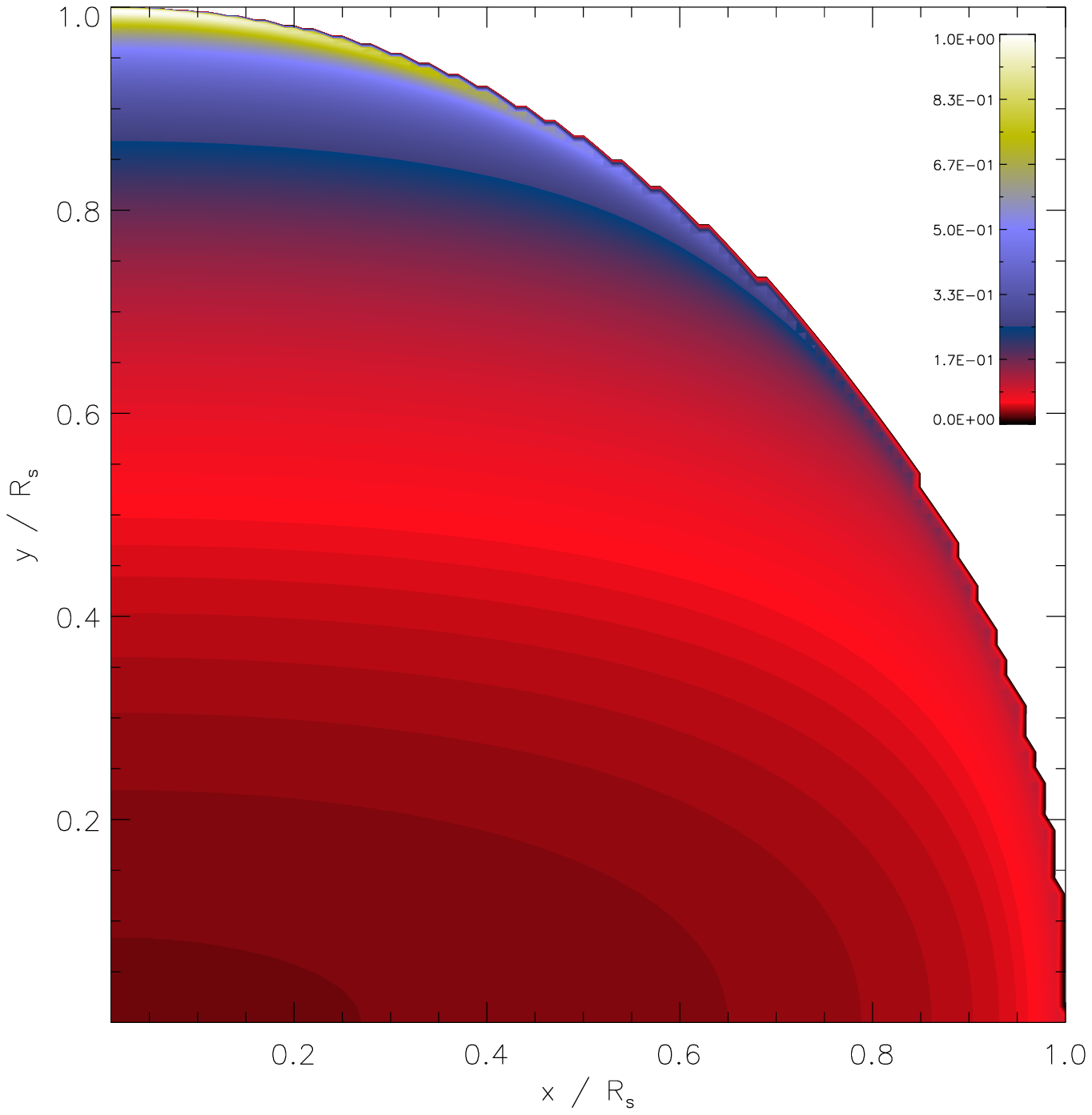} 
\includegraphics[width=6cm]{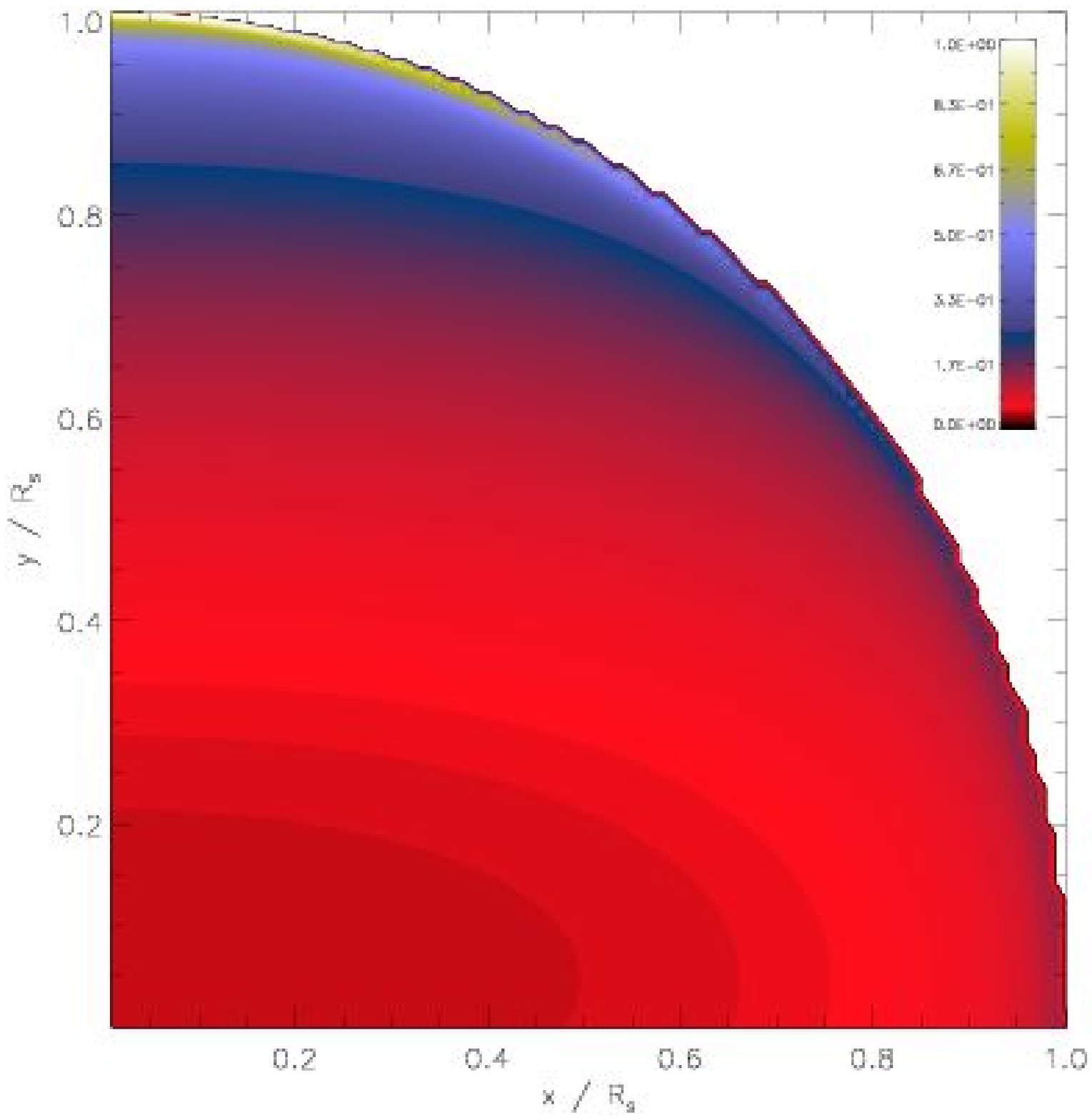} 
\caption{Projected morphology predicted in the case of polar caps (see
Appendix \ref{app-projection}) with $\Delta \tilde{R}=5 \times 10^{-2}$
(\textit{top-left}), $\Delta \tilde{R} = 10^{-2}$ (\textit{top-right}),
$\Delta \tilde{R} = 5 \times 10^{-3}$ (\textit{bottom-left}), $\Delta \tilde{R}$
linearly varying from $2 \times 10^{-3}$ at the pole to $8 \times
10^{-3}$ at the equator (\textit{bottom-right}).  
The radio and the high-energy X-ray
morphology predicted by the model shown in Figure
\ref{fig_model_k0_var_profiles} correspond roughly here to the
top-right and bottom-right panels, respectively. In
all panels, the intensity scaling is square-root.}
\label{fig_projection} 
\end{figure*}




\end{document}